\documentclass[a4paper,12pt]{article}
\usepackage{amsmath,amssymb}
\usepackage{graphicx}
\usepackage{mathrsfs}
\usepackage{float}
\usepackage{braket}
\usepackage{color}
\newcommand{\be}{\begin{equation}}
\newcommand{\ee}{\end{equation}}
\newcommand{\f}{\frac}
\newcommand{\s}{\sqrt}
\newcommand{\p}{\partial}

\newcommand{\bea}{\begin{eqnarray}}
\newcommand{\eea}{\end{eqnarray}}
\newcommand{\ba}{\begin{align}}
\newcommand{\ea}{\end{align}}
\newcommand{\ep}{\epsilon}

\newcommand{\la}{\langle}
\newcommand{\ra}{\rangle}

 \def\f {\frac}

 \def\no{\nonumber \\}

 \def\la{\langle}
 \def\lb{\rangle}
 \def\ep{\epsilon}

\def\ep{{\epsilon}}

\begin{document}

\begin{titlepage}
%%\thispagestyle{empty}
%\begin{flushright}
%TIFR/TH/15-04
%\end{flushright}
\begin{flushright}
YITP-18-78\\
IPMU18-0126\\
\end{flushright}

\vspace{.4cm}

\begin{center}
\noindent{\textbf{Towards an Entanglement Measure for Mixed States in CFTs \\
Based on
 Relative Entropy}}\\
\vspace{1cm}
Tadashi Takayanagi$^{a,b}$\footnote{takayana@yukawa.kyoto-u.ac.jp},
Tomonori Ugajin$^{c}$\footnote{tomonori.ugajin@oist.jp}, and
Koji Umemoto$^{a}$\footnote{koji.umemoto@yukawa.kyoto-u.ac.jp}

\vspace{.5cm}

{\it
$^{a}$Center for Gravitational Physics, \\
Yukawa Institute for Theoretical Physics (YITP), Kyoto University, \\
Kitashirakawa Oiwakecho, Sakyo-ku, Kyoto 606-8502, Japan\\
\vspace{3mm}
 }
{\it
$^{b}$Kavli Institute for the Physics and Mathematics of the Universe,\\
University of Tokyo, Kashiwano-ha, Kashiwa, Chiba 277-8582, Japan\\
\vspace{3mm}
 }
{\it
 $^{c}$Okinawa Institute of Science and Technology, \\
Tancha, Kunigami gun,  Onna son, Okinawa 1919-1 \\
\vspace{0.2cm}
  }
%\vskip
\end{center}
%\vspace{.5cm}
\begin{abstract}
Relative entropy of entanglement (REE) is an entanglement measure  of bipartite mixed states, defined by the minimum of the relative entropy $S(\rho_{AB}|| \sigma_{AB})$ between a given mixed state $\rho_{AB}$ and an arbitrary separable state  $\sigma_{AB}$. The REE is always bounded by the mutual information $I_{AB}=S(\rho_{AB} || \rho_{A}\otimes \rho_{B})$ because the latter measures not only quantum entanglement but also classical correlations.
In this paper we address the question of to what extent REE can be small compared to the mutual information in conformal field theories (CFTs). For this purpose,  we perturbatively compute the relative entropy between the vacuum reduced density matrix $\rho^{0}_{AB}$ on disjoint subsystems $A \cup B$ and arbitrarily separable state $\sigma_{AB}$  in the limit where two subsystems A and B are well separated, then minimize the relative entropy with respect to the separable states.  We argue that the result highly depends on the spectrum of CFT on the subsystems. When we have a few low energy spectrum of operators as in the case where the subsystems consist of finite number of spins in spin chain models, the REE is considerably smaller than the mutual information.  However in general our perturbative scheme breaks down, and the REE can be as large as the mutual information.
\end{abstract}

\end{titlepage}

\tableofcontents

\hspace{0.3cm}

\newpage

\section{Introduction and Summary}

 Quantum entanglement is one of the central ideas in modern theoretical physics. It does not only play crucial roles in quantum information theory but also has a broader range of applications, from condensed matter physics to string theory.

 When we consider a bipartite pure state $|\Psi\lb_{AB}$, we call the state does not have any quantum entanglement when it is represented by a direct product state $|\Psi_1\lb_A\otimes |\Psi_2\lb_B$. For pure states, the amount of quantum entanglement can correctly be measured by the entanglement entropy (or von Neumann entropy): $S(\rho_A)=S(\rho_B)\equiv -\mbox{tr} [\rho_A\log\rho_A]$, where $\rho_A\equiv \mbox{tr} _B|\Psi\lb\la\Psi|$ is the reduced density matrix. This is because the entanglement entropy essentially counts the number of Bell pairs which can be distilled from a given pure state $|\Psi\lb_{AB}$ by local operations and classical communication (LOCC). In LOCC, we can act quantum operations on $A$ and $B$ separately and allow classical communications between $A$ and $B$ at the same time. It is important that the LOCC procedures, which convert a given state into Bell pairs, are reversible for pure states in an asymptotic sense\footnote{Instead of considering a given state itself, one sometimes discusses the procedures on $n$ copies of the original state $\rho_{AB}^{\otimes n}$ followed by the asymptotic ($n\to \infty$) limit. The argument about LOCC reversibility should be correctly taken into account in this regime.}. Namely, after distilling the Bell pairs, one can reproduce the original pure state by performing LOCC on the given Bell pairs. In general, an amount of entanglement quantified by an appropriate entanglement measure has to be always less than the number of Bell pairs necessary to produce a given state by LOCC, and also to be greater than that of Bell pairs distillable from a given state by LOCC. Thus the reversibility guarantees that there is only one measure of quantum entanglement, namely the entanglement entropy \cite{DonaldHorodeckiRudolph01:TheUniqueness}. Refer to the reviews \cite{Calabrese:2009qy,Casini:2009sr,Nishioka:2009un,Rangamani:2016dms,Nishioka:2018khk} for studies of entanglement entropy in quantum field theories and holography.

Next let us turn to a bipartite mixed state, which is described by a density matrix $\rho_{AB}$.
A mixed state $\sigma_{AB}$ has no entanglement if $\sigma_{AB}$ is separable i.e.
\be
\sigma_{AB}=\sum_a p_a\rho^{a}_{A}\otimes \rho^{a}_B, \label{eq; sepst}
\ee
where $p_a$ are positive coefficients such that $\sum_a p_a=1$ and each of $\rho^{a}_{A,B}$ is
a density matrix, which is hermitian and non-negative operator with the unit trace.
 However, the beautiful story which we find for pure states is missing for mixed states because the LOCC procedures of the conversion between a mixed state and Bell pairs is irreversible in general.
 Nevertheless, we can define an entanglement measure by a quantity which is monotonically decreasing under LOCC with a few more optional properties such as the asymptotic continuity.
 We write an entanglement measure for a given bipartite state $\rho_{AB}$ as $E_{\#}(\rho_{AB})$. Such an entanglement measure is far from unique as is clear from the irreversibility (for entanglement measures of mixed states refer to e.g. \cite{Horodecki:2009zz,GOQ} for excellent reviews).

 So far, few calculations of genuine entanglement measures for mixed states have been performed for quantum field theories. The main reasons for this is that the known entanglement measures, such as
  the entanglement of formation $E_F$, the relative entropy of entanglement $E_R$ and the squashed entanglement $E_{Sq}$, all involve very complicated minimization procedures. A correlation measure for mixed state, called entanglement of purification \cite{2002JMP....43.4286T}, involves a slightly simpler minimization procedure, though it is not an entanglement measure. Recently a holographic dual of this quantity has been proposed in \cite{Takayanagi:2017knl,Nguyen:2017yqw} and computations of this quantity in field theories and spin chains have been performed in \cite{Nguyen:2017yqw,Bhattacharyya:2018sbw} (for more progresses refer to \cite{Bao:2017nhh,Blanco:2018riw,Hirai:2018jwy,Espindola:2018ozt,Bao:2018gck,Nomura:2018kji,Umemoto:2018jpc}).
 There is another interesting quantity called the logarithmic negativity \cite{2008PhRvL.101k0501V}, which does not need any minimizations and thus has been successfully computed in two dimensional CFTs \cite{Calabrese:2012ew,Calabrese:2012nk,Calabrese:2013mi}. Though this quantity is monotone under LOCC, the asymptotic continuity condition and convexity are not satisfied. Thus it does not coincide with the entanglement entropy $S(\rho_A)$ when the system $AB$ is pure.

 The main purpose of this paper is to initiate calculations of a true entanglement measure for mixed state in conformal field theories (CFTs). In particular, we focus on the relative entropy of entanglement $E_R(\rho_{AB})$ \cite{Vedral:1997qn, Vedral:1997hd} among entanglement measures, motivated by recent progresses of computational techniques in CFTs of relative entropies \cite{Sarosi:2016oks,Lashkari:2016idm,Sarosi:2016atx,Ugajin:2016opf,Faulkner:2017tkh,Sarosi:2017rsq}. Several bounds for REE in quantum field theories have been obtained in \cite{Hollands:2017dov,Hollands:2017mlk} via an algebraic quantum field theory approach\footnote{
In \cite{Hollands:2017dov},  an  upper bound of $ E_{R} (\rho_{AB})$ in CFT is given: $E_{R} (\rho_{AB}) \leq  N_O  \left(\frac{l}{R}\right)^{2\Delta_{O}}$,
 where $\Delta_O$ is the conformal dimension of lightest primary operator (except the identity) and
$N_O$ is its degeneracy. This follows from Thm 14, Remark 5 of \cite{Hollands:2017dov}. Note that
 when $l/R<<1$, we can approximate $r/R$ in (235) in \cite{Hollands:2017dov} by our $(l/R)^2$ via a conformal transformation \cite{Nakaguchi:2014pha}. Our result in this paper is consistent with this bound and is actually stronger because the REE is at least bounded by the mutual information as in (\ref{boundaree}).} (refer to \cite{Witten:2018zxz} for an excellent review).

The relative entropy of entanglement (REE) is defined as follows. We can measure a distance between two density matrices $\rho$ and $\sigma$ by the relative entropy:
\be
S(\rho||\sigma) ={\rm tr} \;\rho \log \rho - {\rm tr}\; \rho \log \sigma \label{eq:relaa}.
\ee
A basic property of the relative entropy is
$S(\rho ||\sigma)\geq 0$, where the equality holds iff $\rho=\sigma$.

The REE is defined as the shortest distance in the sense of the relative entropy between a given bipartite state $\rho_{AB}$ and an arbitrary separable state $\sigma_{AB}$ as follows:
\be
E_{R}(\rho_{AB}) ={\displaystyle {\inf}_{\sigma_{AB}\in \mbox{Sep}}}  \; \;  S(\rho_{AB} || \sigma_{AB}),\label{eq:REEdef}
\ee
where Sep denotes all separable states. It is obvious that $E_{R}(\rho_{AB})=0$ iff
$\rho_{AB}$ is separable. Moreover, when $\rho_{AB}$ is pure, $E_R(\rho_{AB})$ coincides with the entanglement entropy $S(\rho_A)$.

In this paper we will study the REE $E_{R}$ for  the vacuum reduced density matrix $\rho^0_{AB}$ of CFTs on two disjoint subsystems $A \cup B(\equiv AB)$  in any dimensions. This REE quantifies how much two subsystems $A$ and $B$ are quantum mechanically entangled in a CFT vacuum. We will analyse the REE assuming the subsystems $A$ and $B$ are far apart in terms of power series of
$l/R\ll 1$, where $l$ is the size of $A$ and $B$, while $R$ is the geometrical distance between $A$ and $B$. 

Another useful measure of correlations between $A$ and $B$ is the mutual information:
\be
I(\rho_{AB})=S(\rho_A)+S(\rho_B)-S(\rho_{AB})=S(\rho_{AB}||\rho_A\otimes \rho_B). \label{muta}
\ee
Obviously from the definition of REE, we have the inequality 
\be
E_{R}(\rho_{AB})\leq I(\rho_{AB}).  \label{boundaree}
\ee
This upper bound can also be intuitively understood because the REE measures the amount of quantum entanglement, while the mutual information measures not only quantum entanglement but also
classical correlations. When $A$ and $B$ are far apart, the mutual information for a CFT vacuum (its reduced density matrix is written as $\rho^0_{AB}$) is approximated by the square of vacuum two point function $\la O_{A}O_{B} \ra$ of the (non-trivial) primary operator $O$ with the lowest conformal dimension $\Delta$ (regardless to the positions of operators or the shapes of subsystems):
\be
I(\rho^0_{AB}) \simeq (l)^{4\Delta}  \f{\Gamma (\f{3}{2}) \Gamma(2\Delta +1)}{2\Gamma(2\Delta +\f{3}{2})} \la O_{A}O_{B} \ra^2 \equiv  a_{2\Delta} \left(\f{l}{R} \right)^{4\Delta}. \label{eq:amallmut}
 \ee
For example, the free massless Dirac fermion CFT in two dimensions corresponds to $\Delta=1/2$.
Thus in our limit $l/R\ll 1$, the REE is at least as small as $(l/R)^{4\Delta}$, as can be seen from its upper bound (\ref{boundaree}). Below we are interested in whether the REE can be much smaller than $(l/R)^{4\Delta}$.
 
For general mixed states $\rho$ and $\sigma$, if $\rho-\sigma$ is very small, the relative entropy becomes symmetric $S(\sigma||\rho)\simeq S(\rho || \sigma)$. Therefore, we will first calculate the relative entropy  $S(\sigma_{AB} || \rho_{AB})$ for arbitrary separable density matrices $\sigma_{AB}$, and then take the infinitum with respect to the ensemble $\{ p_{a}, \rho^{a}_{A}, \rho^{a}_{B} \}$. The necessary ingredients for the calculation have been obtained in the previous paper \cite{Ugajin:2016opf} written by the one of the authors, including the vacuum modular Hamiltonian  $K_{AB} = -\log \rho_{AB}$ as well as the von Neumann entropy  $S(\sigma_{AB})$ for any separable density matrices, assuming $l/R\ll 1$.

%In this paper we will show that there is always a separable density matrix $\sigma_{AB}$ whose relative entropy $S(\sigma_{AB} || \rho^0_{AB})$ is vanishing at the order of
%$\left(\f{l}{R} \right)^{4\Delta}$. At least, this shows that
%\be
%E_{R}(\rho^0_{AB}) \ll I(\rho^0_{AB}),  \label{ERE}
%\ee
%when $A$ and $B$ are enough far apart $l/R\ll 1$. Since the mutual information (\ref{eq:amallmut}) measures both classical and quantum correlations, our result for the REE (\ref{ERE}), compared with (\ref{eq:amallmut}), tells us that the correlations between the subsystem $A$ and $B$ are essentially classical when $A$ and $B$ are far apart. Moreover, under the assumption that the lightest primary is always dominant, we will show that $S(\sigma_{AB}|| \rho^0_{AB})$ for a certain separable state $\sigma_{AB}$, is vanishing at each order of perturbative expansion with respect to $\left(\frac{l}{R}\right)\ll 1$ and this strongly suggests that the REE $E_{R}(\rho^0_{AB})$ also vanishes in the same way.

In this paper we first compute the contribution of the lightest primary operator to the relative entropy, then minimizing it by assuming it gives the dominant contribution in the large separation limit, as in  case of the mutual information. We are able to show that we can make this contribution always vanish by appropriately choosing the separable state at any order of the perturbation.  
We also give an explanation why the separable state is indistinguishable from $\rho^{0}_{AB}$ from the viewpoint of local observables.

However, the minimization becomes much more complicated when we include the effects of other operators with higher conformal dimensions. In this case, we find that our perturbative calculation is not enough, since we cannot suppress the expectation value of higher dimensional operators in general.

From these observations we argue that the behavior of REE is highly dependent on the operator spectrum of CFT in the subsystems. For a CFT with few low energy states such as the case where the subsystems consist of finite number of spins in spin chain models, the perturbative analysis is enough and we find that there is tiny quantum entanglement as $E_{R}(\rho^0_{AB}) \ll I(\rho^0_{AB})$. 
We can check this statement by  having  an independent argument in spin chain models.
 
However, in generic setups our perturbative expansion gets uncontrollable and this implies that
the REE can be as large as the mutual information $I_{AB}$. Especially we expect $E_{R}(\rho^0_{AB}) \simeq I(\rho^0_{AB})$ for holographic CFTs, as the operator spectrum does not seem to allow us to optimize the minimizations in the definition of REE. On the other hand, since integrable CFTs such as the rational CFTs in two dimensions, have simple operator spectrum and algebra, there might be a chance that the REE can be smaller than the mutual information even when the subsystems are much 
larger than the lattice spacing. For further investigations, we probably need to develop methods which does not rely on perturbations.

The organization of this paper is as follows:  In section \ref{section;REER}, we review
basic properties of the relative entropy of entanglement. In section \ref {section;RelativeinCFT}, after explaining the basic set up,  we compute the relative entropy $S(\sigma_{AB} || \rho^0_{AB}) $ between the vacuum reduced density matrix $\rho^0_{AB}$ and an arbitrary separable state $\sigma_{AB}$ in the leading order of the large distance limit $l/R \rightarrow 0$, based on results of \cite{Ugajin:2016opf}. In section \ref{sec:Minimization}, we minimize the relative entropy with respect to the separable states.   We find there alway be a separable state whose relative entropy is vanishing therefore $E_{R}( \rho^0_{AB} ) =0$ at the quadratic order of perturbative expansions.
In section \ref{sec;nextleading} we take into account of higher order perturbative corrections, and argue they do not change our result under certain conditions. In section \ref{nextlp}, we 
discuss the contribution from the next lightest primary, which shows the result of REE is very sensitive to the operator spectrum. In section \ref{discus}, we will compare our results with other known results and discuss future problems. In the appendix we explain the details of our calculations.

\section{Properties of Relative Entropy of Entanglement}

\label{section;REER}

The relative entropy of entanglement $E_{R}(\rho_{AB})$ is defined by (\ref{eq:REEdef}) for a bipartite quantum state $\rho_{AB}$,
i.e. the shortest distance between $\rho_{AB}$ and the set of separable
states measured by the relative entropy. 

\subsection{Properties of REE}

The properties of REE is summarized as follows (for more details, refer to \cite{Horodecki:2009zz,GOQ})\footnote{In this section we deal with the finite dimensional Hilbert space for simplicity. Most of the properties and the inequalities are also proven in the infinite dimensional setup, refer to \cite{Hollands:2017dov, Duan2017} for recent discussion.}:\\

(i) Faithfulness: $E_{R}(\rho_{AB})\geq0$ and $E_{R}(\rho_{AB})=0$ if and only
if $\rho_{AB}$ is separable.\\

(ii) Monotonicity: $E_{R}(\rho_{AB})$ is monotonically decreasing under (stochastic)
LOCC.\\

(iii) Convexity:  $E_{R}(\rho_{AB})$ is convex i.e. $E_{R}(x\rho_{AB}+(1-x)\rho'_{AB})\leq xE_{R}(\rho_{AB})+(1-x)E_{R}(\rho'_{AB})$
for any $x\in[0,1]$.\\

(iv) Continuity: $E_{R}(\rho_{AB})$ is continuous respect to $\rho_{AB}$ i.e. if $\rho_{AB}$ and $\sigma_{AB}$ are close in trace distance, then the value of $E_{R}(\rho_{AB})$ approaches that of $E_{R}(\sigma_{AB})$ \footnote{There are many variations of the continuity of entanglement measures. In particular, REE is also asymptotic continuous, which is described by the limit of many copies $\lim_{n\to\infty}\rho_{AB}^{\otimes n}$ and an important property in the axiomatic approach of entanglement measures.} :
\be
||\rho_{AB}-\sigma_{AB}||\to0,\ {\rm then}\ \frac{|E_R(\rho_{AB})-E_R(\sigma_{AB})|}{\log \dim \mathcal{H}_{AB}}\to0,
\ee
where $\mathcal{H}_{AB}$ is the Hilbert space $\rho_{AB}$ and $\sigma_{AB}$ act on \cite{DonaldHorodecki99:Continuity}.\\

(v) Subadditivity: $E_{R}(\rho_{AB})$ always satisfies the subadditivity $E_{R}(\rho_{AB}\otimes\rho'_{A'B'}) \leq E_{R}(\rho_{AB})+E_{R}(\rho_{A'B'}')$. Note that it does not satisfy the additivity $E_{R}(\rho_{AB}\otimes\rho_{A'B'}')= E_{R}(\rho_{AB})+E_{R}(\rho_{A'B'}')$ in general.\\

(vi) When $\rho_{AB}$ is pure, $E_{R}(\rho_{AB})$ reduces to the
entanglement entropy $S(\rho_{A})(=S(\rho_{B}))$.  To see this, consider a pure state $\rho_{AB}=\ket{\psi}\bra{\psi}_{AB}$ with the Schmidt decomposition
\be
\ket{\psi}_{AB} = \sum_{i} \sqrt{\lambda_i} \ket{i}_A\ket{i}_B,
\ee
where $\lambda_i\geq0,\ \sum_i \lambda_i=1$. Then it is shown that the closest separable state of $\rho_{AB}$ which reaches the minimization in (\ref{eq:REEdef}) is given by a simple form \cite{Vedral:1997hd,0305-4470-33-22-101}

\be
\sigma_{AB} = \sum_{i} \lambda_i \ket{i}\bra{i}_A\otimes\ket{i}\bra{i}_B.
\ee
Indeed, one can easily check that $S(\rho_{AB}||\sigma_{AB})$ of these states reduces to the entanglement entropy:
\be
S(\rho_{AB}||\sigma_{AB})=-{\rm tr}\rho_{AB}\log\sigma_{AB} = -\sum_{i} \lambda_i \log \lambda_i = S(\rho_A).
\ee
\\
Above properties indicate that REE is a good generalization of entanglement entropy to a genuine entanglement measure for mixed states.

There are several upper/lower bounds for REE: As we have already mentioned, $E_{R}(\rho_{AB})$ is bounded from above by the mutual information
$I(\rho_{AB})=S(\rho_{AB}||\rho_{A}\otimes\rho_{B})$ as $E_{R}(\rho_{AB})\leq I(\rho_{AB})$,
which follows directly from the definition of REE. Another upper bound is given the entanglement of formation $E_R(\rho_{AB})\leq E_F(\rho_{AB})$, which is also a good measure of entanglement for mixed states. On the other hand, a lower bound
is given by the distillable entanglement $E_{D}(\rho_{AB})\leq E_{R}(\rho_{AB})$, which counts the number of EPR pairs extractable from a given state $\rho_{AB}$ by LOCC.  This bound also leads to an entropic inequality $E_R(\rho_{AB})\geq\max[S(\rho_A),S(\rho_B)]-S(\rho_{AB})$ \footnote{This inequality can be rewritten in terms of conditional entropy $S(B|A)=S(\rho_{AB})-S(\rho_A)$ as $E_R(\rho_{AB})\geq \max[-S(A|B),-S(B|A)]$, which was firstly derived in \cite{0305-4470-33-22-101}.} by virtue of the hashing inequality \cite{Devetak207}.  It may also be worth noting that there is no generic inequality relationship between REE and the negativity \cite{1464-4266-6-12-009}. 

%restriction to the marginal separable states or reversed components \cite{0305-4470-36-20-316}, and

%(Pinsker's inequality) by $S(\rho||\sigma)\geq\frac{1}{2}\mbox{tr}[\sqrt[]{(\rho-\sigma)^{2}}]$.

\subsection{Quadratic Approximations}
In the present paper we will deal with $S(\sigma_{AB}||\rho_{AB})$ rather
than $S(\rho_{AB}||\sigma_{AB})$ for technical simplicity, where $\sigma_{AB}$ represents a separable state. This does not change the main results at the quadratic order of small perturbation of quantum state. Consider the case
where $\rho$ and $\sigma$ are very closed to each other
\begin{equation}
\rho=\sigma+\delta\rho.
\end{equation}
If we expand $S(\rho||\sigma)$ up to the quadratic order of $\delta\rho$,
we find (see e.g.\cite{Faulkner:2017tkh})
\begin{equation}
S(\rho||\sigma)=\frac{1}{2}{\rm tr}\left[\delta\rho\left.\frac{d}{dx}\log(\sigma+x\delta\rho)\right|_{x=0}\right]+O(\delta\rho^{3}).
\end{equation}
From this expression, it is clear that $S(\rho||\sigma)$ coincides
with the reversed one $S(\sigma||\rho)$ up to the quadratic order
\begin{equation}
S(\rho||\sigma)-S(\sigma||\rho)=O(\delta\rho^{3}).\label{equivree}
\end{equation}
One can also understand this symmetry as a consequence from positivity and non-degeneracy of the
relative entropy.

As an illustration, consider the case where $\sigma$ and $\rho$
are $2\times2$ density matrices, expressed as:
\begin{equation}
\sigma=\begin{pmatrix}\alpha & 0\\
0 & 1-\alpha
\end{pmatrix},\ \rho=\begin{pmatrix}\alpha+\epsilon & \delta_{1}+i\delta_{2}\\
\delta_{1}-i\delta_{2} & 1-\alpha-\epsilon
\end{pmatrix},
\end{equation}
and treat $\delta_{1}$ and $\delta_{2}$ as infinitesimally small
real parameters. We require $0<\alpha<1$ for positivity of density matrix. If we only keep up to quadratic
terms of them, we can confirm the equivalence (\ref{equivree}) explicitly
as follows:
\begin{equation}
S(\rho||\sigma)=S(\sigma||\rho)=\frac{\epsilon^{2}}{2\alpha(1-\alpha)}+\frac{\log\frac{1-\alpha}{\alpha}}{1-2\alpha}(\delta_{1}^{2}+\delta_{2}^{2}).
\end{equation}

In \cite{0305-4470-36-20-316}, an entanglement measure so-called the reversed REE was introduced in the same spirit of REE with reversed components:
\be
E_{RR}(\rho_{AB}) ={\displaystyle {\inf}_{\sigma_{AB}\in \mbox{Sep,LI}}}  \; \;  S(\sigma_{AB} || \rho_{AB}) \label{eq:revREEdef},
\ee
where the minimization is restricted to a class of separable states locally identical to $\rho_{AB}$ i.e. ${\rm tr}_B(\sigma_{AB})=\rho_A,\ {\rm tr}_A(\sigma_{AB}) = \rho_B$. This quantity also satisfies many properties of a good entanglement measure, especially the additivity. However, when $\rho_{AB}$ is pure, $E_{RR}(\rho_{AB})$ generically diverges (or trivially vanishes) and thus it can not be regarded as an appropriate generalization of entanglement entropy for mixed states.

\section{The Calculation of the Relative Entropy} \label{section;RelativeinCFT}
\subsection{Set up}

%The purpose of this note is evaluating the relative entropy in CFT  in the limit where  the two regions are well  separated. In this limit the mutual information $I_{AB}$ between   region A and region   B is  given by the two point  function of the lightest operator $O$,
%\be
%I_{AB} =(l_{A})^{2\Delta} (l_{B})^{2\Delta} \f{\Gamma (\f{3}{2}) \Gamma(2\Delta +1)}{2\Gamma(2\Delta +\f{3}{2})} \la O_{A}O_{B} \ra^2,\label{eq:amallmut}
% \ee
%
%where $\Delta$ is the conformal dimension of $O$.

We begin with  a CFT on a $d$ dimensional flat space $\mathbb{R}^d$, and two ball shaped regions $A$ and $B$, with the radius $l$ and the distance $R$.  In this section we estimate the relative entropy $S(\sigma_{AB}|| \rho^0_{AB})$  between  the vacuum reduced density matrix on $A \cup B$ defined by,
\be
\rho^0_{AB} = {\rm tr}_{(AB)^{c}} |0 \ra \la 0|
\ee
and an arbitrary separable density matrix $\sigma_{AB}$, in the large distance limit $l/ R \rightarrow  0$.
\footnote{Precisely speaking, in the actual computation we regard this set up as a particular limit of the system on a cylinder $ \mathbb{R} \times S^{d-1} $. Let $L$ be the radius of the spacial sphere $S^{d-1} $, then the large distance limit in  $\mathbb{R}$ is equivalent to the double scaling limit on the cylinder,
\be
\f{l}{L} \rightarrow 0, \quad \f{l}{R} \rightarrow 0.
\ee
 }

It is convenient to split the relative entropy into two parts:
\be
 S(\sigma_{AB}|| \rho^0_{AB}) =-S(\sigma_{AB}) +{\rm tr} \; \sigma_{AB} K^0_{AB},
\ee

where $S(\sigma_{AB})$  is the von Neumann  entropy of  the separable density matrix and $K_{AB}$ is the modular Hamiltonian of $\rho^0_{AB}$,
\be
K^0_{AB} =-\log \rho^0_{AB}.
\ee

\begin{figure}[t]
\centering
  \includegraphics[width=10cm]{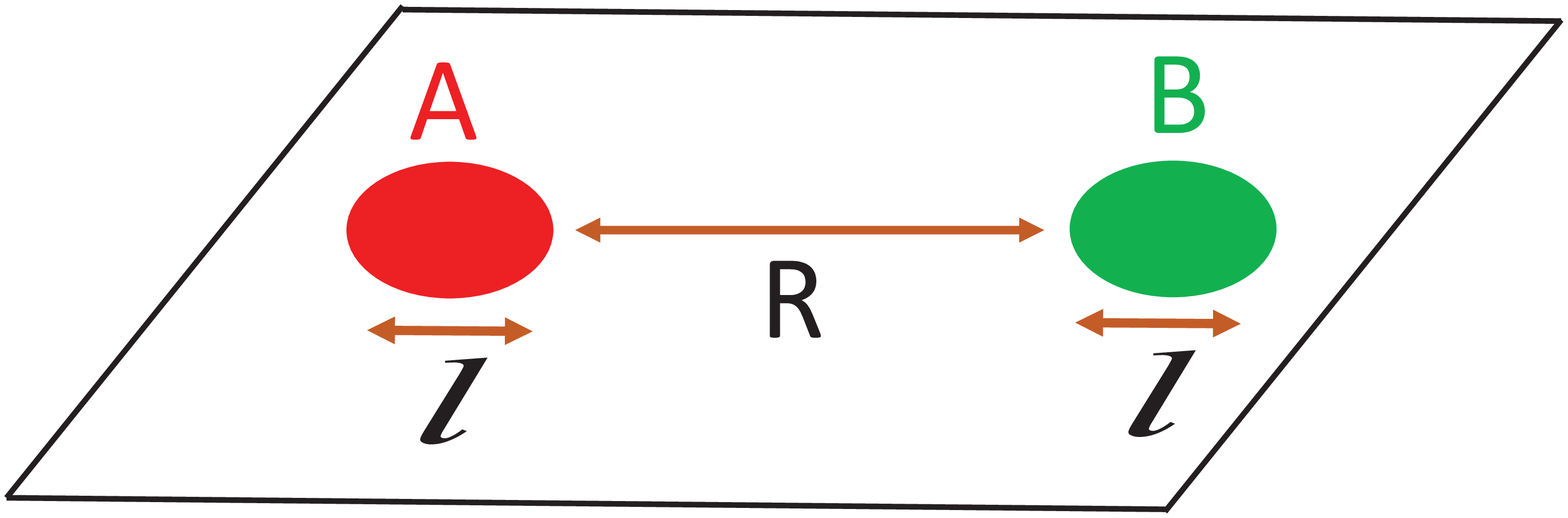}
\caption{The choice of subsystem $A$ and $B$ to define the REE $E_R(\rho_{AB})$.}
\end{figure}

\subsection{The calculation of  $S(\sigma_{AB})$} \label{subsection;sepentro}

In this subsection  we explain how to compute the von Neumann entropy, $S(\sigma_{AB})$ for a separable state
$\sigma_{AB}$. This is  a slight generalization of the previous calculation done in \cite{Sarosi:2016atx, Ugajin:2016opf}.  Here we only outline the calculation, and leave details in appendix A.

For this purpose, we employ the usual replica trick,
\be
S(\sigma_{AB})  =\lim_{n \rightarrow 1} \f{1}{1-n} \log {\rm tr} \; \sigma_{AB}^{n}.
\ee

This R\'enyi entropy   can be expanded as
\begin{align}
 {\rm tr} \; \sigma_{AB}^{n} & =\sum_{\{a_{k}\}} \prod^{n-1}_{k=0} p_{a_{k}}  {\rm tr} \left[ \left( \rho^{a_{1}}_{A} \otimes \rho^{a_{1}}_{B}\right)  \cdots \left( \rho^{a_{n}}_{A} \otimes \rho^{a_{n}}_{B}\right) \right] \nonumber  \\
&= \sum_{\{a_{k}\}} \prod^{n-1}_{k=0}\;  p_{a_{k}}  {\rm tr} \left[  \rho^{a_{1}}_{A} \cdots  \rho^{a_{n}}_{A} \right]
 {\rm tr} \left[ \rho^{a_{1}}_{B} \cdots  \rho^{a_{n}}_{B} \right].  \label{eq: renyiee}
\end{align}

We first compute the right hand side of (\ref{eq: renyiee}) for reduced density matrices of global excitations, $|X_{a} \ra, \;|Y_{a} \ra $ ($a=0$ corresponds to the vacuum: $|X_0\ra=|Y_0\ra=|0\ra$)

\be
\rho^{a}_{A}  ={\rm tr}_{A^{c}} | X_{a} \ra \la X_{a} |, \quad  \rho^{a}_{B} ={\rm tr}_{B^{c}} | Y_{a} \ra \la Y_{a} |,
\label{eq:two density}
\ee

on cylinder  $\mathbb{R} \times S^{d-1}$ with the metric,
\be
ds^{2} =dt^2 +d\theta^2 + \sin^{2} \theta d\Omega_{d-2}^2.
\ee

We then  read off the  result for arbitrary $\rho^{a}_{A}, \rho^{a}_{B}$ from it.    We take both subsystems $A,B$ to be  isomorphic to the ball shaped region on the spatial sphere $S^{d-1}$,
\be
A,B: [0, l/2] \times S^{d-2}.
\ee
Also it is important to notice that in this calculation we do not need to specify the distance between  two
regions.

State operator correspondence allows us to write the quantities in the right hand side  in terms of the 2n point  correlation functions  on the covering space  $\Sigma_{n}= S^{1}_{n} \times H^{d-1}$  \cite{Sarosi:2016atx},
\be
{\rm tr}  [\rho^{a_{1}}_{A} \cdots  \rho^{a_{n}}_{A}] = \f{\la\prod_{k=0}^{n-1} X_{a_{k}}(w_{k}) X_{a_{k}} (\hat{w}_{k}) \ra_{\Sigma_{n}}}{\prod_{k=0}^{n-1} \la X_{a_{k}}(w_{0}) X_{a_{k}} (\hat{w}_{0}) \ra_{\Sigma_{1}} }\cdot \frac{Z^{(n)}_A}{(Z^{(1)}_A)^n}, \label{eq:corr}
\ee
where $X_{a_{k}} (w_{k})$ is the local operator corresponding to the global state $ | X_{a_{k}} \ra$
and there is a similar relation for the subsystem $B$ and the global state $ | Y_{a_{k}} \ra$; also
$Z^{(n)}_A$ denotes the vacuum partition function on $\Sigma_n$. The correlation functions are 
normalized such that $\la 1\lb_{\Sigma_n}=1$.

The covering space $\Sigma_{n}$ is equipped with the metric,
\be
ds^{2}_{\Sigma_{n}} =d\tau^2 +du^2+ \sinh^2 u d\Omega_{d-2}^2,  \quad \tau \sim \tau +2\pi n,
\ee

and the locations of the local operators are given by
\be
w_{k}: (\tau_{k}, u_{k}) = \left(2\pi( k+\f{1}{2}) +\f{l}{2},0 \right), \quad \hat{w}_{k}: (\tau_{k}, u_{k}) =\left ( 2\pi( k+\f{1}{2})-\f{l}{2} ,0 \right). \label{eq;loccover}
\ee

The small subsystem size limit $l \rightarrow0$  corresponds to choose  the particular channel  $w_{k} \rightarrow \hat{w}_{k}$  of these correlation functions. There one can expand them by OPE. By picking up the contribution of the lightest primary operator $O$ with the conformal dimension $\Delta$. By taking the analytic continuation $n \rightarrow 1$ of the R\'enyi entropy,
we finally obtain\footnote{ We choose the $a=0$ component to be  reduced density matrices of the vacuum, ie \be
\rho^{0}_{A} ={\rm tr}_{\bar{A}} |0 \ra \la 0|,  \quad  \rho^{0}_{B} ={\rm tr}_{\bar{B}} |0 \ra \la 0|
\ee
}

\begin{align}
-S( \sigma_{AB})&=
 -\sum_{a} p_{a} \left( \la K^{0}_{A} \rho^{a}_{A} \ra+ \la K^{0}_{B} \rho^{a}_{B} \ra \right) \nonumber \\
&+a_{\Delta}\left(l\right)^{2\Delta} \left[\left(\sum_{a} p_{a} \la \rho^{a}_{A}O \ra\right)^2+\left(\sum_{a} p_{a} \la \rho^{a}_{B}O \ra \right)^2\right] \nonumber \\
&-C_{OOO}b_{\Delta}l^{3\Delta} \left[\left(\sum_{a} p_{a} \la \rho^{a}_{A}O \ra\right)^3+\left(\sum_{a} p_{a} \la \rho^{a}_{B}O \ra \right)^3 \right] \nonumber \\
&+ a_{2\Delta}\left(l\right)^{4\Delta}\left[ \sum_{a}p_{a}  \la \rho^{a}_{A} O_{A} \ra \la \rho^{a}_{B} O_{B}\ra -  \left(\sum_{a} p_{a} \la \rho^{a}_{A}O \ra\right) \left(\sum_{a} p_{a} \la \rho^{a}_{B}O \ra \right)\right]^2 \label{eq:sepent}
\end{align}

where $K^{0}_{A}$ is the vacuum modular Hamiltonian on the region $A$. In a CFT vacuum on a ball shaped region, $K^{0}_{A}$ is given by a simple integral of stress tensor.  We do not need its precise form, as
it is always canceled with other contributions in the relative entropies.

Meanwhile, the von Neumann entropy of a reduced density matrix $\rho_{A}$   on the single subsystem $A$  is given by (see for example  \cite{Sarosi:2016atx} )
\be
S(\rho_{A})  = {\rm  tr} \left[ K^{0}_{A} \rho_{A} \right] -a_{\Delta} l_{A}^{2\Delta}\;  {\rm tr} \left[ \rho_{A} O\right]^2 + C_{OOO} b_{\Delta} (l_{A})^{3\Delta}  {\rm tr} \left[ \rho_{A} O\right]^3 +\cdots \label{eq: eeformula}
\ee

with
\be
a_{\Delta}= \f{\Gamma (\f{3}{2}) \Gamma(\Delta +1)}{2\Gamma(\Delta +\f{3}{2})}, \quad b_{\Delta} =\f{2\s{\pi}}{3\Gamma (\f{3\Delta+3}{2})},
\ee

and $C_{OOO}$ is the OPE coefficient of the primary $O$.

Our result indicates the von Neumann entropy of $\sigma_{AB}$ gets factorized
\be
S(\sigma_{AB})  = S(\sigma_{A})+ S(\sigma_{B}), \quad  \sigma_{A} =\sum_{a} p_{a} \rho^{a}_{A},\; \sigma_{B} =\sum_{a} p_{a} \rho^{a}_{B}
\ee
 up to $l^{3\Delta}$ order, and the effect of the classical
correlation first enters at $l^{4\Delta}$ order.  If we write the correlation part in terms of original separable density matrix $\sigma_{AB}$
\be
S(\sigma_{A})+S(\sigma_{B})-S(\sigma_{AB})=a_{2\Delta}\left(l\right)^{4\Delta} \left[\la  \sigma_{AB} O_{A} O_{B} \ra - \la \sigma_{A} O_{A} \ra \la \sigma_{B} O_{B} \ra  \right]^2,
\ee
therefore this part is basically the square of the connected part of the two point function $\la O_{A} O_{B} \ra$ evaluated on $\sigma_{AB}$.

This can be compared with the mutual information $I_{AB} (\rho^0_{AB})$ of a reduced density matrix  $\rho^0_{AB}$ at this $l^{4\Delta}$ order \cite{Ugajin:2016opf},
\be
I_{AB} (\rho^0_{AB})  =a_{2\Delta}\left(l\right)^{4\Delta} \left[\la  \rho_{AB} O_{A} O_{B} \ra - \la \rho_{A} O_{A} \ra \la \rho_{B} O_{B} \ra  \right]^2=a_{2\Delta}\left(\frac{l}{R}\right)^{4\Delta}, \label{eq:muinf}
\ee
and the two results are related by the exchange $\sigma_{AB} \leftrightarrow \rho^0_{AB}$.
Indeed, as is clear from the  discussion in the appendix B,   the derivations of the two results are identical to each other, once we identify the two correlation functions $\la \sigma_{AB} O_{A} O_{B} \ra  \leftrightarrow   \la \rho^0_{AB} O_{A} O_{B} \ra$.

\subsection{Modular Hamiltonian and Calculation of ${\rm tr}\; \sigma_{AB} K^{0}_{AB}$}

Having calculated the von Neumann entropy part,  let us move on to the modular Hamiltonian part,
\be
{\rm tr} \; \sigma_{AB} K^{0}_{AB}, \quad K^{0}_{AB} =-\log \rho^0_{AB}.
\ee

It was shown in \cite{Ugajin:2016opf}, $K_{AB}$ takes following form,
\be
K^{0}_{AB} =K^{0}_{A}+K^{0}_{B} + \tilde{K}^{0}_{AB},
\ee
and in the large distance limit $\f{l}{R} \rightarrow 0$, we have
\be
\tilde{K}^{0}_{AB}  =-2a_{2\Delta}\;  l^{4\Delta} \la O_{A} O_{B} \ra O_{A} O_{B} + I_{AB}.\label{eq:modab}
\ee
This was obtained by starting from the expression of von Neumann entropy $S(\rho_{AB})$ for a generic state $\rho_{AB}$ which is  related to the mutual information (\ref{eq:muinf}), and applying the ``first law trick'', which will be reviewed in section 5. More details of the discussion can be again found in \cite{Ugajin:2016opf}. $I_{AB}$ in   (\ref{eq:modab}) denotes the constant part of the modular Hamiltonian. We need this part in order to make sure the relation
\be
S_{AB} =\la \rho_{AB} K_{AB} \ra
\ee
and  $I_{AB}$ coincides with the value of the vacuum mutual information (\ref{eq:amallmut}).  Then,

\begin{align}
{\rm tr} \left[\sigma_{AB} K^{0}_{AB} \right]& =\sum_{a} p_{a} \left[ \la \rho^{a}_{A}K^{0}_{A} \ra+\la \rho_{B}^{a} K^{0}_{B} \ra \right]  \nonumber \\
&-  2 a_{2\Delta}  \left(\f{l_{A}}{R}\right)^{4\Delta} \sum_{a} p_{a} \left[ \la \rho^{a}_{A} O_{A} \ra  \la \rho^{a}_{B}  O_{B} \ra  \right] + I_{AB}.  \label{eq:mutexp}
\end{align}

\subsection{Net result}

Combining  (\ref{eq:sepent}) (\ref{eq:mutexp}),  the relative entropy we would like to minimize is

\begin{align}
S(\sigma_{AB}||\rho_{AB})&=a_{\Delta}\left(l\right)^{2\Delta} \left[\left(\sum_{a} p_{a} \la \rho^{a}_{A}O \ra\right)^2+\left(\sum_{a} p_{a} \la \rho^{a}_{B}O \ra \right)^2\right] \nonumber \\
&-C_{OOO}b_{\Delta}l^{3\Delta} \left[\left(\sum_{a} p_{a} \la \rho^{a}_{A}O \ra\right)^3+\left(\sum_{a} p_{a} \la \rho^{a}_{B}O \ra \right)^3 \right] \nonumber \\
&+ a_{2\Delta}\left(l\right)^{4\Delta}\left[ \sum_{a}p_{a}  \la \rho^{a}_{A} O_{A} \ra \la \rho^{a}_{B} O_{B}\ra -  \left(\sum_{a} p_{a} \la \rho^{a}_{A}O \ra\right) \left(\sum_{a} p_{a} \la \rho^{a}_{B}O \ra \right)\right]^2 \nonumber \\
&-  2 a_{2\Delta} (l)^{2\Delta} \left(\f{l}{R}\right)^{2\Delta} \sum_{a} p_{a} \left[ \la \rho^{a}_{A} O_{A} \ra  \la \rho^{a}_{B}  O_{B} \ra  \right] + I_{AB}.   \label{eq:relativegen}
\end{align}

Notice that there are higher order corrections. We will discuss on this in section \ref{sec;nextleading}.

\section{Minimization}  \label{sec:Minimization}

In the previous section we computed the relative entropy $S(\sigma_{AB}||\rho^0_{AB})$ between the vacuum reduced density matrix and an arbitrary separable density matrix $\sigma_{AB}$ in the large distance limit $\f{l}{R} \rightarrow 0$ keeping only the contributions from the lightest primary operator. In this section, we would like to find the separable density matrix that minimizes the relative entropy and compute the relative entropy of entanglement $E_{R}(\rho^0_{AB})$.
We choose the separable state $\sigma_{AB}$ to be in the form:
\be
\sigma_{AB} = (1-\varepsilon) \rho^0_{A} \otimes \rho^0_{B}  + \varepsilon \rho^{1}_{A} \otimes   \rho^{1}_{B} ,
\ee
where $\ep$ is a small parameter and $\rho^0_{A}=\mbox{tr}_B\rho^0_{AB}$. In addition, $\rho^1_{A,B}$ are arbitrary density matrices with non-vanishing one-point function of the primary $O$, which is defined to be
\be
\mbox{tr}[\rho^1_A O_A]=\mbox{tr}[\rho^1_B O_B]=l^{-\Delta}x,  \ \ \ (x>0).   \label{opffr}
\ee

We would like to keep only quadratic perturbations to $S(\sigma_{AB}||\rho^0_{AB})$ so that we have
$S(\sigma_{AB}||\rho^0_{AB})\simeq S(\rho^0_{AB}||\sigma_{AB})$ as in (\ref{equivree}). To implement this, we
define the small perturbations $\delta\rho^0$ and $\delta\rho^1$ by
\be
\delta\rho^0=\rho^0_{AB}-\rho^0_A\otimes \rho^0_B,\ \ \
\delta\rho^1=\ep(\rho^1_A\otimes\rho^1_B-\rho^0_A\otimes \rho^0_B),
\ee
such that
\be
\rho^0_{AB}-\sigma_{AB}=\delta\rho^0-\delta\rho^1.  \label{pertrho}
\ee

Our perturbations are parameterized by the following two small parameters:
\bea
&& W\equiv l^{2\Delta}\mbox{tr}[\delta\rho^0 O_AO_B]=l^{2\Delta}\la O_AO_B\lb=\left(\frac{l}{R}\right)^{2\Delta}\ll 1,\no
&& Z\equiv l^{2\Delta}\mbox{tr}[\delta\rho^1 O_AO_B]=l^{2\Delta}\sum_a p_a(\mbox{tr}\rho^a_AO_A)(\mbox{tr}\rho^a_BO_B)=\ep x^2\ll 1.
 \eea

It will be useful to note that the mutual information (\ref{eq:muinf}) when $A$ and $B$ are far apart
is at the quadratic order. Indeed, we have
\be
I(\rho^0_{AB})=S(\rho^0_{AB}|\rho^0_A\otimes \rho^0_B)\simeq S(\rho^0_A\otimes \rho^0_B|\rho^0_{AB})
\simeq a_{2\Delta}W^2.
\ee

In this parametrization, our result in the small interval expansion (\ref{eq:relativegen}) is expresses as follows up to the quadratic order of $Z$ and $W$:
\be
S(\sigma_{AB}||\rho^0_{AB})= \left(a_{2\Delta}+\frac{2a_\Delta}{x^2}\right)Z^2-2a_{2\Delta}ZW+a_{2\Delta}W^2.
\label{eq: relativefin}
\ee
By varying $Z$ (or equally $\ep$) to  minimize the relative entropy, we obtain
\be
\mbox{Min}_{Z}\left[S(\sigma_{AB}||\rho^0_{AB})\right]=\left(\frac{2a_{\Delta}a_{2\Delta}}{2a_{\Delta}+a_{2\Delta}x^2}
\right)W^2,
\ee
at $Z=\frac{a_{2\Delta}x^2}{2a_{\Delta}+a_{2\Delta}x^2}W$.

Next we vary the choice of the state $\rho^1_{A,B}$ so that the one-point function (\ref{opffr})
gets larger such that $Z=\ep x^2$ is still very small. It is obvious that we can define such a state with
an arbitrary large $x$ in the continuous limit of field theories. In the limit,
\be
x\to \infty, \ \  \ep\to 0,\ \   \mbox{with}\ \ \  \ep x^2 \simeq \left(\frac{l}{R}\right)^{2\Delta}\ll 1,
\ee
we find that the infimum of the relative entropy is vanishing
\bea
\mbox{inf}_{Z,x}\left[S(\sigma_{AB}||\rho^0_{AB})\right]=0,
\eea
up to the quadratic order.
Note that at this infimum, the separable state is locally vacuum on the region $A$ and $B$, i.e. ${\rm tr}_{A,B} \sigma_{AB} ={\rm tr}_{A,B} \rho_{AB}$.

Finally, by employing the relation (\ref{equivree}) up to the quadratic order of our perturbation
(\ref{pertrho}), we obtain the estimation of REE:
\be
E_{R}(\rho^0_{AB}) =0\cdot \left(\frac{l}{R}\right)^{4\Delta}+\mbox{higher orders of $(l/R)$}. \label{resulta}
\ee
This manifestly shows that the REE is much smaller than the mutual information
\be
\frac{E_{R}(\rho^0_{AB})}{I(\rho^0_{AB})} \to 0,  \label{resultb}
\ee
in the limit $(l/R)\to 0$ where $A$ and $B$ are far apart. However, notice again that in this calculation we only keep contributions from the lightest primary operator.

\subsection{An Interpretation}\label{subsection;corr}
There is an intuitive way to understand why the separable density matrix $\sigma_{AB}$ is indistinguishable from the vacuum reduced density matrix $\rho^{0}_{AB}$.

It is useful to write the separable density matrix,
\be
\sigma_{AB} =\lim_{x \rightarrow \infty} \left[ \left(1 -\f{l^{2\Delta}\la O_{A} O_{B} \ra}{x^{2}} \right) \rho^{0}_{A} \rho^{0}_{B} + \f{l^{2\Delta}\la O_{A} O_{B} \ra}{x^{2}} \rho^{1}_{A} \rho^{1}_{B}\right]. \label{sepminnn}
\ee

Notice  that this separable density matrix  $\sigma_{AB}$  reproduces all correlation functions of $\rho_{AB}$  on the disjoint  region $ A \cup B$,  as it should be. In our small subsystem limit, 
if we truncate the spectrum to the lightest primary operator,  we only need to reproduce one and two point functions of $\{ 1,O\}$:
\be
{\rm tr} \left[ \rho^{0}_{AB} O_{A} O_{B} \right], \quad {\rm tr} \left[ \rho^{0}_{A} O_{A} \right] =  {\rm tr} \left[ \rho^{0}_{B} O_{B} \right] =0.
\ee
 We can easily see that this is indeed the case,
\be
{\rm tr} \left[ \rho^{0}_{AB} O_{A} O_{B} \right] ={\rm tr} \left[ \sigma_{AB} O_{A} O_{B} \right], \quad \left[ \sigma_{A} O_{A} \right] =  {\rm tr} \left[ \sigma_{B} O_{B} \right] =0.
\ee
As we will see in the final section, this result corresponds to a critical spin chain example where 
the subsystem $A$ and $B$ consist of finite number of spins.

Furthermore, this observation makes it clear that for $m$ disjoint subsystems $A_{1}\cup \cdots A_{m}$ the separable density matrix which minimize the analogous relative entropy is given by
\be
\sigma_{A_{1}, \cdots A_{m}} =\left(1-\sum_{k=1}^{m} \sum_{\{i_{1} \cdots i_{k} \}} P^{(k)}_{\{i_{1} \cdots i_{k} \}} \right) \rho^{0}_{A_{1}} \cdots \rho^{0}_{A_{n}}   + \sum_{k=1}^{m} \sum_{\{i_{1} \cdots i_{k} \}} P^{(k)}_{\{i_{1} \cdots i_{k} \}} \rho_{i_{1}, i_{2} \cdots i_{k}}
\ee
with

\be
P^{(k)}_{\{i_{1} \cdots i_{k} \}} = \lim_{x \rightarrow \infty} \f{l^{k\Delta}\la O_{A_{i_{1}}} O_{A_{i_{2}}} \cdots O_{A_{i_{n}}} \ra}{x^{k}}, \quad \rho_{i_{1} \cdots i_{k}} = \rho^{0}_{A_{1}} \cdots \rho^{1}_{A_{i_{1}}} \cdots \rho^{1}_{A_{i_{k}}}\cdots
\rho^{0}_{A_{i_{n}}}.
\ee

One can easily see that the density matrix reproduce all k  $( \leq m)  $point functions of $O$

\subsection{An example of the separable density matrix $\sigma_{AB}$ in 2d CFT}

One can indeed construct a one parameter family of density matrices $\{ \rho_{\beta} \}$, $\beta \rightarrow 0$ of which realizes the  infimum in a class of  two dimensional conformal field theory.  Suppose that the lightest primary operator of the  2d CFT in question is the stress tensor $O=T_{zz}$. The we can  take $\rho^{1}_{A} $ defined by

\be
\rho^{1}_{A}  ={\rm tr}_{A^{c}} | \psi_{\beta} \ra \la \psi_{\beta}|, \quad  \psi_{\beta}  =\f{e^{-\beta H} }{\s{N}}|B \ra
\ee
where $| B \ra$ is a boundary state of the CFT, and $N$ is the normalization factor.
Then its stress tensor expectation value  is
\be
x_{\beta}  =l^2 \la \psi_{\beta}| T_{zz} | \psi_{\beta} \ra = \f{cl^2}{24 \beta^2},
\ee
and $x_{\beta} \rightarrow \infty$ when $\beta \rightarrow 0$.

This implies that if we define $\rho_{\beta}$ by
\be
\rho_{\beta} = (1- \varepsilon_{0}(x_{\beta})) \rho^{0}_{A} \otimes \rho^{0}_{B} + \varepsilon_{0}(x_{\beta}) \rho^{1}_{A} \otimes  \rho^{1}_{B},
\ee

then  the density matrix,
\be
\sigma_{AB} \equiv \lim_{\beta \rightarrow 0} \rho_{\beta}
\ee
 is indistinguishable from the vacuum reduced density matrix $ \rho^0_{AB}$, at least in the  $\left(\f{l}{R} \right)^{8}$ order.

If we consider a discretized lattice model such as spin chains and introduce the lattice spacing $a$,
then the minimum possible value of the parameter $\beta$ is $O(a)$. In more general, we expect that for a generic operator with the dimension $\Delta$, the maximal value of $x$ will behave like 
\be
x_{max}\sim \left(\frac{l}{a}\right)^{\Delta}.  \label{maxx}
\ee

\section{Next Leading Order} \label{sec;nextleading}

In the previous section we found the relative entropy of entanglement $E_{R} (\rho^{0}_{AB})$ is vanishing up to $\left( \f{l}{R}\right)^{4\Delta}$ order.  It is natural to ask whether higher order corrections can modify this result or not. Motivated by this question, in this section we compute $S(\sigma_{AB}|| \rho^0_{AB})$ up to $\left( \f{l}{R}\right)^{6\Delta}$ by again assuming the lightest primary plays still a dominant role at this order. We also use the fact that the one point functions of  the separable state   $\sigma_{AB}$ must be vanishing,
\be
{\rm tr} \left[\sigma_{AB} O_{A} \right] ={\rm tr} \left[\sigma_{AB} O_{B} \right] =0, \label{eq;locvac1}
\ee
in order to reproduce the vacuum one point functions. Restricting $\sigma_{AB}$ to be in this class of states
drastically simplifies the computation below.  Notice that  from (\ref{eq: eeformula}) this in particular  implies that
\be
S(\sum_{a} p_{a} \rho^{a}_{A} )=S(\rho^{0}_{A}), \quad  S(\sum_{a} p_{a} \rho^{a}_{B} )=S(\rho^{0}_{B}).  \label{eq;locvac2}
\ee

\subsection{$S(\sigma_{AB})$}

 The von Neumann entropy $S(\sigma_{AB}) $ can be computed along the line of section \ref{subsection;sepentro} by further expanding the correlator (\ref{eq: renyiee}),  in particular  allowing 3 $O$ s to propagate in the internal lines of it.  The final result of the cubic order is given by (see appendix A for more details):

\be
S (\sigma_{AB}) \Big|_{l^{6\Delta}}=  (l)^{6\Delta} \left(\sum_{a} p_{a} \la \rho^{a} O \ra^2 \right)^3  C_{OOO}^2 \f{\Gamma(\f{1+2\Delta}{2})^3}{12\pi \Gamma(\f{3+6\Delta}{2})},
\ee
and we can write
\be
S (\sigma_{AB}) \Big|_{l^{6\Delta}} = \left( d_{\Delta} C_{OOO}^2 \right) Z^{3} ,\label{eq:sbb}
\ee
where  $d_{\Delta} \equiv 2^{6\Delta} \f{\Gamma(\f{1+2\Delta}{2})^3}{12\pi \Gamma(\f{3+6\Delta}{2})}$.

\subsection{${\rm tr} \; \sigma_{AB} K^{0}_{AB} $}

Next let us compute the expectation value of the modular Hamiltonian at this order.  First of all, the von Neumann entropy of a reduced density matrix $\rho_{AB}$ satisfying the locally vacuum condition  (\ref{eq;locvac1}), (\ref{eq;locvac2}) (but not necessary a separable state) is directly related to its mutual information,
\be
 S(\rho_{AB}) =S(\rho^{0}_{A})+S(\rho^{0}_{B}) -I_{AB}(\rho_{AB}),
\ee
where $\rho^{0}_{A,B}$ is the vacuum reduced density matrix on the region $A,B$ respectively.

This mutual information can be computed either directly by a correlator  with twist operators in the replica trick   or indirectly from $S(\sigma_{AB})$ by the replacement in (\ref{eq:sbb})\footnote{For the detail of this replacement, see Appendix \ref{sec;replacement}.}
\be
\quad I_{AB} (\rho_{AB}) \Big|_{l^{6\Delta}} = -\left( d_{\Delta} C_{OOO}^2 \right) W(\rho_{AB})^{3},
\ee
where $W(\rho_{AB})=l^{2\Delta}\mbox{tr}[\rho_{AB}O_AO_B]$.

We can use this expression of mutual information for $ \rho_{AB}$ satisfying the locally vacuum condition to read off the form of vacuum modular Hamiltonian $K^{0}_{AB}$ at $l^{6\Delta}$ order, by using the
first law trick. Imagine starting from the vacuum reduced density matrix $ \rho^{0}_{AB}$ , and slightly deform it
$ \rho^{0}_{AB}\rightarrow \rho_{AB}= \rho^{0}_{AB}+ \delta \rho^{0}_{AB}$,  then
the value of mutual information $I(\rho_{AB})$ as well as entanglement entropy $I(\rho_{AB})$ are changed by the deformation. In particular the first order change satisfies the first law. If we know the form of $S(\rho_{AB})$ for any $\rho_{AB}$, we can read off the form of modular Hamiltonian from the above equation.  In our current case it goes like,
\begin{align}
\delta S\Big|_{l^{6\Delta}}  =-\delta I_{AB}\Big|_{l^{6\Delta}}  = +3\left( d_{\Delta} C_{OOO}^2 \right)  W^2 {\rm tr} \left[ \delta \rho_{AB} O_{A}O_{B}\right],
\end{align}
with $W=W(\rho^{0}_{AB})$.
Since this is true for any $\delta \rho_{AB}$ satisfying the locally vacuum condition, we derive the form of modular Hamiltonian at this order
\be
K^{0}_{AB} \Big|_{l^{6\Delta}} = +3\left( d_{\Delta} C_{OOO}^2 \right)  W(\rho^{0}_{AB})^2 O_{A}O_{B} +a_{AB},
\ee

where $a_{AB}$ is the constant part of the modular Hamiltonian, fixed by the relation $S(\rho^{0}_{AB}) = {\rm tr} \left[\rho^{0}_{AB} K^{0}_{AB} \right]$. In this case,
\be
a_{AB} = -2\left( d_{\Delta} C_{OOO}^2 \right) W^3.
\ee

By plugging these expressions, we get
\be
{\rm tr} \; \sigma_{AB} K^{0}_{AB}\Big|_{l^{6\Delta}}  = (d_{\Delta}C_{OOO}^2 )(3ZW^2-2W^3)
\ee

Again notice that  the form of $S(\rho_{AB})$ is not generic, and valid only when $\rho_{AB}$ satisfies the locally vacuum condition.  Therefore
 the form of modular Hamiltonian we derive from the expression  is only true when it is acted on the space of reduced density matrix satisfying the condition.
 However it is sufficient  for our purpose of computing the expectation value of vacuum modular Hamiltonian with respect to a separable $\sigma_{AB}$ which satisfies the condition.

A more rigorous argument is as follows. Again consider  the change of the density matrix   $ \rho^{0}_{AB}\rightarrow \sigma_{AB}= \rho^{0}_{AB}+ \delta \rho^{0}_{AB}$, then

\begin{align}
\delta S_{AB}&\equiv {\rm tr} \left[ K^{0}_{AB}(\sigma_{AB}-\rho^0_{AB}) \right] +O(\delta \rho^2) \nonumber \\
&= 3 (l)^{6\Delta} {\rm tr} \left[(\sigma_{AB}-\rho^0_{AB})  O_{A} O_{B} \right]   W^{2} (d_{\Delta}C_{OOO}^2 ) +O(\delta \rho^2),\nonumber \\
&=3d_{\Delta}C_{OOO}^2W^2(Z-W) +O(\delta \rho^2).
\end{align}

From this we can read off the value which we want as follows
\begin{align}
{\rm tr} \; \sigma_{AB} K^{0}_{AB} &=  {\rm tr} \left[ K^{0}_{AB}(\sigma_{AB}-\rho^0_{AB}) \right] + {\rm tr} \;  \rho^0_{AB} K^{0}_{AB} \nonumber \\[+10 pt]
&=  (d_{\Delta}C_{OOO}^2 )(3ZW^2-2W^3) \label{eq;6thmdodh}
\end{align}

in the derivation we do not need to use the precise form of the modular Hamiltonian.

%\be
%\la K^{0}_{AB} \delta \rho  \ra = -3(d_{\Delta}C_{OOO}^2 ) W^2 Z \label{eq:6thmodh}
%\ee
%
%If we take

%
%  The $ (l)^{6\Delta}$ of the mutual information of a locally vacuum state is again computed by replica trick ,
%and just given by with the replacement $J_{OO} \rightarrow \la \rho_{AB} O_{A} O_{B} \ra$. We
%\be
%I_{AB}\Big|_{l^{6\Delta}} =- (l)^{6\Delta} \la \rho_{AB} O_{A} O_{B} \ra^3  (d_{\Delta}C_{OOO}^2 )
%\ee
%
%We write the vacuum mutual information $I^{0}_{AB}$ by defining
%\be
%W\equiv l^{2\Delta} \la \rho^{0}_{AB} O_{A} O_{B} \ra,
%\ee
%
%then
%\be
%I^{0}_{AB}\Big|_{l^{6\Delta}} = - \left( d_{\Delta} C_{OOO}^2 \right) W^{3} \label{eq:6thmut}

%\subsection{ Modular Hamiltonian part }
%
% because of the locally vacuum condition, now
%\be
%S_{AB} =-I_{AB}.
%\ee
%
%If we slightly deform the density matrix from the vacuum one $\rho^{(0)}_{AB}$ with keeping the condition intact, we have

%\textcolor{blue}{ We need to evaluate the constant term of the modular Hamiltonian,  the result is
%\be
%a_{AB}=a_{2\Delta} l^{4\Delta} \la O_{A} O_{B} \ra^2 -2 (d_{\Delta} C_{OOO}^2) l^{6\Delta}  \la O_{A} O_{B} \ra^3
%\ee
%which unfortunately modifies the discussion below.}
%
%Since now ,
%\be
%\delta \rho = \sum_{a} p_{a} \;\rho^{a}_{A}\; \rho^{a}_{B} -\rho^{0}_{AB},
%\ee

%we have
%\be
%\la K^{0}_{AB} \delta \rho  \ra = -3(d_{\Delta}C_{OOO}^2 ) W^2 Z \label{eq:6thmodh}
%\ee

\subsection{Minimization}

Combining these results, (\ref{eq:sbb}) and (\ref{eq;6thmdodh}),
we obtain the expression of relative entropy up to this order $l^{6\Delta}$
\begin{align}
S(\sigma_{AB}|| \rho^0_{AB})  &=a_{2\Delta } (W^2-2WZ+Z^2) \nonumber \\
&-d_{\Delta} C_{OOO}^2 (2W^3-3W^2Z+ Z^3).
\end{align}
This function again has a minima at $Z=W$, where $S(\sigma_{AB}|| \rho^0_{AB})$ is vanishing.

 One may worry that  this relative entropy negatively diverges in $Z \rightarrow \infty$ limit. Of course this is just an artifact of our truncation  the perturbative expansion, and the local minima $Z=W$ should be the global minima, as is clear from the argument found in section \ref{subsection;corr}.

As long as we assume that only the primary operator $O$ is relevant, the above argument of vanishing
$S(\sigma_{AB}|| \rho^0_{AB})$ at $Z=W$ continues to be true  in all orders in the perturbative expansion with respect to $Z$ and $W$.  First, in this expansion the von Neumann entropy $S(\sigma_{AB})$  is expressed as
\be
S(\sigma_{AB}) = \sum_{n} b_{n}  Z^{n} ,\label{eq;sepst}
\ee
where $b_{n} $ s are unknown coefficients depending on $\Delta$ and $C_{OOO}$, though we do not need their precise values in the argument below.  The modular Hamiltonian
expectation value ${\rm tr} \; \sigma_{AB} K^{0}_{AB} $ can again be read off from the mutual information of locally vacuum state, which is related to  (\ref{eq;sepst}) by replacing  $Z$ to the corresponding two point function,
\be
{\rm tr} \; \sigma_{AB} K^{0}_{AB} =\sum_{n}  b_{n} \left[ nW^{n-1} Z- (n-1)W^{n} \right]
\ee

Finally the relative entropy is given by
\be
S(\sigma_{AB}|| \rho_{AB}) = -\sum_{n}  b_{n} \left[Z^{n} -nW^{n-1} Z+(n-1) W^{n} \right] .
\ee
By taking derivative with respect to $Z$, we see that  each term in the expansion has the minimum at $W=Z$ where the relative entropy vanishes.

In this section we have shown that under the assumption that the primary $O$, which as the lowest conformal dimension, gives dominant contributions in each order of $\left(\frac{l}{R}\right)$ expansions, the minimum of relative entropy $S(\sigma_{AB}|| \rho^0_{AB})$ vanishes. Even though we cannot use the relation (\ref{equivree}) for perturbations higher than quadratic order, the vanishing relative entropy shows that the
vacuum reduced density matrix $\rho^0_{AB}$ is very closed to the separable states at each order of perturbation. Therefore our result here suggests that the reversed one $S(\rho^0_{AB}||\sigma_{AB})$ and
the REE $E_{R}(\rho^0_{AB})$ vanishes in each perturbative order.

\section{Contribution from the Next Lightest Primary}\label{nextlp}

So far, we have been discussing possible higher order corrections due to the exchanges of  the lightest primary operator. There is another type of corrections to the relative entropy, which is coming from exchanges of heavier operators.  To get some intuitions for this, here we study the  effect of the next lightest primary  $O_{NL}$ with the conformal dimension $\Delta_{NL}$.

If we assume the locally vacuum condition, the contribution of   $O_{NL}$ to the relative entropy first enters at $l^{2\Delta+2\Delta_{NL}}$ order.
From the replica calculation we find the expression of $S(\sigma_{AB})$, up to this order,

\be
 -S(\sigma_{AB}) =a_{2\Delta} Z^{2} +2a_{(\Delta+\Delta_{NL})} Z_{1}^2, \quad Z_{1} \equiv l^{\Delta+\Delta_{NL}}\sum_{a} p_{a} \la \rho^{a}_{A} O  \ra  \la \rho^{a}_{A} O_{NL}  \ra .
\ee

Similarly the mutual information of generic $\rho_{AB}$ up to this order is
\be
I_{AB}(\rho_{AB} ) = a_{2\Delta} W(\rho_{AB})^2 +a_{(\Delta+\Delta_{NL})}   l^{2(\Delta+\Delta_{NL})} \left({\rm tr} [ \rho_{AB} O_{A} O_{B,NL}] + {\rm tr} [ \rho_{AB} O_{B} O_{A,NL}] \right)
\ee
Notice however the second term vanishes once we set $\rho_{AB}= \rho^{0}_{AB}$ thus the modular Hamiltonian part does not receive correction at this order.

The net result of the relative entropy up to this order is therefore
\be
S(\sigma_{AB}|| \rho^{0}_{AB}) =  a_{2\Delta} (W-Z)^2 + 2a_{(\Delta+\Delta_{NL})} Z_{1}^2 . \label{eq;NLresult}
\ee

We then minimize this relative entropy.   If we can regard  second term of (\ref{eq;NLresult}) as a perturbative correction to the first term of order $l^{2\Delta_{NL}}$, then the first order correction to the minimum value of the relative entropy is evaluated just by substituting the separable density matrix (\ref{sepminnn}) that minimizes the relative entropy at the leading order.  The value of $Z_{1}$ for this separable state is given by
\be
Z_{1} = \la O_{A} O_{B} \ra \f{x_{NL}}{x} , \quad x_{NL} = l^{\Delta_{NL}}\la \rho^{1} O_{NL} \ra
\ee

In order for this to work,  we need to require $ x\gg x_{NL}$.  However it seems difficult to find such $\rho_{1}$ in general especially when we need to take $x$ to be large.  If we naively construct such $\rho_{1}$ with large $x$, we fail. This is because the maximal value of $x$ and $x_{NL}$ scales as in (\ref{maxx}) in terms of the lattice spacing $a$: $x\sim (l/a)^\Delta$ and $x_{NL}\sim (l/a)^{\Delta_{NL}}$. Thus we generically expect $x_{NL}\gg x$, assuming $l\gg a$.

From the above analysis of the contribution from the next lightest operator,  
 it does not seem to be possible to reduce the relative entropy 
$S(\sigma_{AB}||\rho^0_{AB})$ in generic CFTs, by fine-tuning the separable state $\sigma_{AB}$ as far as we assume our perturbative analysis.

\section{Conclusions and Discussions}\label{discus}

In this paper, we considered the relative entropy of entanglement (REE) $E_{R}(\rho^0_{AB})$ for CFT vacua. We focus on the case where the subsystem $A$ and $B$ are largely separated compared with their sizes. In this limit we can employ the OPE expansions in terms of operators localized in $A$ and $B$.

\subsection{Lightest Operator Dominant Case and Spin Chain Example} 

In the first part of this paper, we assumed that the lightest primary operator gives the dominant contribution. Under this assumption we were able to show that $E_{R}(\rho^0_{AB})$ gets much smaller than the mutual information $I(\rho_{AB})$ as in (\ref{resulta}) and (\ref{resultb}). This means that the vacuum reduced density matrix $\rho_{AB}$ is an almost separable state. Moreover, under the assumption that the lightest primary is always dominant, we showed that $S(\sigma_{AB}|| \rho^0_{AB})$ for a certain separable state $\sigma_{AB}$, is vanishing at each order of power expansions of $\left(\frac{l}{R}\right)$ and this strongly suggests that the REE $E_{R}(\rho^0_{AB})$
also vanishes in the same way. Thus we find that the correlations between $A$ and $B$ are classical 
in this case.

We expect that the assumption of taking into account only the lightest primary can be justified when 
we consider a critical spin chain model and the subsystems consist of finite numbers of spins. For this, let us consider a $S=1/2$ spin chain at a quantum critical point and choose the subsystem $A$ and $B$ to be
the $p$-th and $(p+R)$-th spin, denoted by $\sigma^A_i$ and $\sigma^B_i$, where $i=1,2,3$ i.e. the Pauli matrices, which satisfy the relation $\mbox{Tr}[\sigma_i\sigma_j]=2\delta_{ij}$.
The correlation function looks like
\be
\la \sigma^A_i\sigma^B_j\lb\simeq \delta_{ij}|R|^{-2\Delta}\equiv \gamma\cdot \delta_{ij}.
\label{twopt}
\ee
where $\Delta$ is the dimension of the spin operator. Note that when the distance $R$ between two spins are large the magnitude $\gamma$ gets very small.

In this setup, the reduced density matrix for $AB$ is given by
\bea
\rho_{AB}=\frac{I_{AB}}{4}+\frac{\gamma}{4}\sum^{3}_{i=1}\left(\sigma^A_i\otimes \sigma^B_i\right).
\eea
In the $4\times 4$ matrix form this reads
\bea
\rho_{AB}=\left(\begin{array}{cccc}
  1+\gamma & 0 & 0 & 0 \\
  0 & 1-\gamma & 2\gamma & 0 \\
  0 & 2\gamma &1-\gamma & 0 \\
  0 & 0 & 0 & 1+\gamma \\
\end{array}\right).
\eea
The requirement of positivity of density matrix is expresses as $-1<\gamma<1/3$. If $\gamma$ is small as we consider, this condition is clearly satisfied.

Since the dimension $H_A\otimes H_B$ is less than six, we know that the condition of separability is equivalent to the PPT criterion (positivity under partial transposition) \cite{Sep}. The density matrix under the partial transposition (transposition w.r.t $B$) reads
\bea
(\rho_{AB})^{T_B}=\left(\begin{array}{cccc}
  1+\gamma & 0 & 0 & 2\gamma \\
  0 & 1-\gamma & 0 & 0 \\
  0 & 0 &1-\gamma & 0 \\
  2\gamma & 0 & 0 & 1+\gamma \\
\end{array}\right).
\eea
In this case the PPT criterion says that $\rho_{AB}$ is separable
if and only if  $-\frac{1}{3}<\gamma<1$.

In summary $\rho_{AB}$ is separable when $-1/3<\gamma<1/3$ and is not separable
(i.e. is entangled) when $-1<\gamma<-1/3$. Thus, in our spin chain example, when the distance
$R$ between $A$ and $B$ are large (i.e. $\gamma$ is very small), we can conclude that $\rho_{AB}$ is separable and the logarithmic negativity defined by ${\mathcal{E}}=\log |(\rho_{AB})^{T_B}|$ is vanishing, where $T_B$ is transposition only for $B$ (called partial transposition).

For a larger spin $S\geq 1$, or for larger subsystems $A$ and $B$, the PPT criterion and separability are not equivalent. However, still it is known that the state (in a finite dimensional Hilbert space) which is very closed to the maximally mixed state $\rho=\frac{I_N}{N}$ is separable
\cite{SepV}. Therefore if two spins are far apart and their correlation functions are small,
we can apply this theorem to find that $\rho_{AB}$ is separable. 

Indeed, the above results for spin chains are consistent with our field theoretic result that the REE is vanishing in our perturbation theory.

\subsection{Generic Cases and Holographic CFTs}

In the later part of this paper, we estimated the contribution from the next lightest primary.
This analysis tells us that the higher dimensional operators can give substantial contributions 
to the relative entropy in general, which violates our perturbation theory. The main reason for this is that if we want to choose a state $\rho^1$ with a very large expectation value of the lightest primary, then the expectation value of a heavier operator for the same state also inevitably gets larger.

For example, if we consider holographic CFTs, the lightest primary is typically a single trace operator. The double trace operator has the contribution $x_{double}=x_{single}^2$ and thus cannot be negligible. This suggests that in holographic CFT, we have $E_R(\rho^0_{AB})\simeq I_{AB}(\rho^0_{AB})$, i.e. the correlations between $A$ and $B$ origin from  quantum entanglement.\footnote{
 The analysis of holographic entanglement entropy
\cite{Ryu:2006bv} shows that the holographic mutual information satisfies the monogamy as shown in \cite{Hayden:2011ag}. This suggests that the leading order part $O(N^2)$ (i.e. classical gravity part) of holographic entanglement entropy originates from quantum entanglement. In our analysis we take the large separation limit between $A$ and $B$ and thus such a classical gravity contribution is vanishing. Thus, in this paper, we are interested in the higher order part $O(1)$, which is dual to quantum effects in gravity.}

Computations of the REEs for integrable CFTs, such as rational CFTs in two dimensions, will need careful treatments. Interestingly, in \cite{Calabrese:2012ew, Calabrese:2012nk,Calabrese:2013mi}, the logarithmic negativity in the same setup as ours was computed in two dimensional CFTs and spin chains and was shown to be much smaller than any powers of $l/R$ for rational CFTs. The logarithmic negativity is known to be monotone under LOCC and is vanishing for all separable states, though can be zero even for non-separable states. In this sense, the relation between the REE and logarithmic negativity is not straightforward. However, this result strongly implies that the quantum entanglement is highly reduced. In our analysis of REE, since the primary operator spectrum and its OPE algebra are simple, it might be possible that the argument for generic CFTs in the above cannot be applied. If so, the REE can be smaller. To completely answer this question, we need to develop calculations of relative entropy beyond our perturbation theory, which is an interesting future problem.

\section*{Acknowledgments}
We thank Arpan Bhattacharyya, Pawel Caputa, Horacio Casini, Patrick Hayden, Veronika Hubeny, Yuya Kusuki, Robert Myers,Xiao-liang Qi, Mukund Rangamani, Shinsei Ryu, G\'abor S\'arosi, and  Erik Tonni  for useful discussions. TT would also like to thank Okinawa Institute of Science and Technology Graduate University (OIST) and Stanford Institute for Theoretical Physics for their hospitality where this work was progressed. TT is supported by JSPS Grant-in-Aid for Scientific Research (A) No.16H02182 and by JSPS Grant-in-Aid for Challenging Research (Exploratory) 18K18766. KU is supported by Grant-in-Aid for JSPS Fellows No.18J22888. TT is also supported by the Simons Foundation through the ``It from Qubit'' collaboration and by World Premier International Research Center Initiative (WPI Initiative) from the Japan Ministry of Education, Culture, Sports, Science and Technology (MEXT).

\appendix

\section{Calculation of $S(\sigma_{AB})=S({\displaystyle \sum_{a} } \;p_{a}\;  \rho^{a}_{A} \otimes \rho^{a}_{B})$}

In this section we calculate  $S( \sum_{a}\;  p_{a} \; \rho^{a}_{A} \otimes \rho^{a}_{B})$ perturbatively  in the small subsystem size expansion.

For a moment we consider the density matrices coming  from tracing out  global excited states $|X_{a} \ra, \;|Y_{a} \ra $ on cylinder (\ref{eq:two density}), so that their R\'enyi entropies are computed by (after applying several conformal mappings) the corresponding correlation function on n sheet covering space $\Sigma_{n} =S^{1}_{n} \times H^{d-1} $,
\be
{\rm tr}  \rho^{a_{1}}_{A} \cdots  \rho^{a_{n}}_{A} = \f{\la\prod_{k=0}^{n-1} X_{a_{k}}(w_{k}) X_{a_{k}} (\hat{w}_{k}) \ra_{\Sigma_{n}}}{\prod_{k=0}^{n-1} \la X_{a_{k}}(w_{0}) X_{a_{k}} (\hat{w}_{0}) \ra_{\Sigma_{1}} } \; \cdot\frac{Z^{(n)}_{A}}{(Z^{(1)}_{A})^n},
\ee
where the locations of these operators  $w_{k}, \hat{w}_{k}$ are  defined in (\ref{eq;loccover}).
Note also that the correlation functions are normalized such that $\la 1\lb_{\Sigma_n}=1$.

In the small  subsystem size limit $ 2l \rightarrow 0$, $w_{k} \rightarrow  \hat{w}_{k}$.
Also  we have
\be
Z^{(n)}_{A} ={\rm tr}\;  (\rho^{0}_{A})^{n}, \quad \rho^{0}_{A} ={\rm tr} |0 \ra \la 0|.
\ee

\begin{figure}[t]
\centering
  \includegraphics[width=10cm]{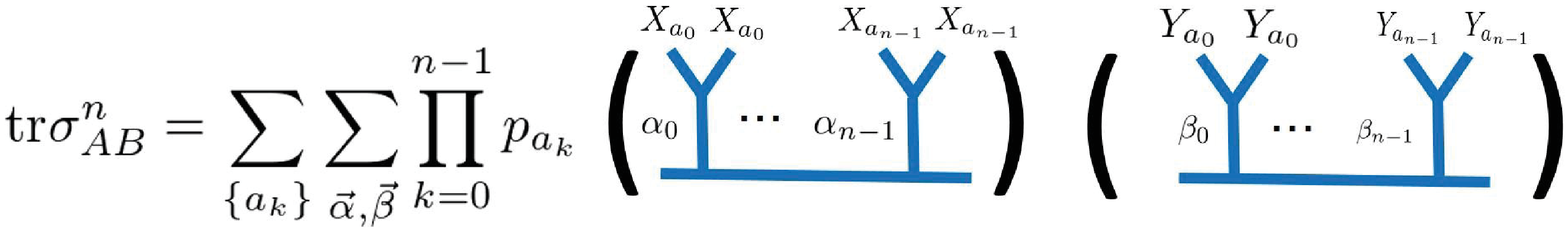}
\caption{A graphical representation of the expansion  the Renyi entropy}
\end{figure}

 From this we have an expression of the R\'enyi entropy in terms of correlation functions,
\begin{align}
{\rm tr} \; \sigma_{AB}^n &= \sum_{\{a_{k}\}} \prod^{n-1}_{k=0} p_{a_{k}}  {\rm tr} \left[ \left( \rho^{a_{1}}_{A} \otimes \rho^{a_{1}}_{B}\right)  \cdots \left( \rho^{a_{n}}_{A} \otimes \rho^{a_{n}}_{B}\right) \right] \nonumber \\
&= \sum_{\{a_{k}\}} \prod^{n-1}_{k=0} p_{a_{k}}  \left(\f{\la\prod_{k=0}^{n-1} X_{a_{k}}(w_{k}) X_{a_{k}} (\hat{w}_{k}) \ra_{\Sigma_{n}}}{\prod_{k=0}^{n-1} \la X_{a_{k}}(w_{k}) X_{a_{k}} (\hat{w}_{k}) \ra_{\Sigma_{1}} } \right) \left(\f{\la\prod_{k=0}^{n-1} Y_{a_{k}}(w'_{k}) Y_{a_{k}} (\hat{w}'_{k}) \ra_{\Sigma_{n}}}{\prod_{k=0}^{n-1} \la Y_{a_{k}}(w'_{0}) Y_{a_{k}} (\hat{w}'_{0}) \ra_{\Sigma_{1}} } \right)  \frac{Z^{(n)}_{A}Z^{(n)}_{B}}{(Z^{(1)}_{A}Z^{(1)}_B)^n}.  \label{eq: renyicomp}
\end{align}

$(w'_{k}, \hat{w}'_{k}$) are again the locations of  the local operators for the   subsystem B.  The strategy to calculate the right hand side of (\ref{eq: renyicomp}) is as usual, expanding the correlation functions by using
OPEs
\be
\f{X_{a_{k}}(w_{k}) X_{a_{k}}(\hat{w}_{k})}{ \la X_{a_{k}}(w_{0}) X_{a_{k}}(\hat{w}_{0}) \ra_{\Sigma_{1}}} = \f{ \la X_{a_{k}}(w_{k}) X_{a_{k}}(\hat{w}_{k}) \ra_{\Sigma_{n}}}{ \la X_{a_{k}}(w_{0}) X_{a_{k}}(\hat{w}_{0}) \ra_{\Sigma_{1}}} \sum_{\alpha_{k}} C_{X_{a_{k}}X_{a_{k}} \alpha_{k}}\; ( 2l)^{\Delta_{\alpha_{k}}} \alpha_{k} (w_{k})
\ee

where $\alpha_{k}$s are  the operators propagating the internal line, and by $\Delta_{\alpha_{k}}$ we denote the scaling dimension of  $\alpha_{k}$.
We also have similar expansion of $Y$'s
\be
\f{Y_{a_{k}}(w'_{k}) Y_{a_{k}}(\hat{w}'_{k})}{ \la Y_{a_{k}}(w_{0}) Y_{a_{k}}(\hat{w}_{0}) \ra_{\Sigma_{1}}} = \f{ \la Y_{a_{k}}(w'_{k}) Y_{a_{k}}(\hat{w}'_{k}) \ra_{\Sigma_{n}}}{ \la Y_{a_{k}}(w'_{0}) Y_{a_{k}}(\hat{w}'_{0}) \ra_{\Sigma_{1}}} \sum_{\alpha_{k}} C_{Y_{a_{k}}Y_{a_{k}} \beta_{k}}\; ( 2l)^{\Delta_{\beta_{k}}} \beta_{k} (w'_{k}).
\ee

Using these formulae
\bea
&& {\rm tr} \; \sigma_{AB}^n\cdot \frac{(Z^{(1)}_{A} Z^{(1)}_{B})^n}{Z^{(n)}_{A} Z^{(n)}_{B} } \no
&& =\sum_{\{ \alpha_{0}, \cdots \alpha_{n-1} \}, \{ \beta_{0}, \cdots \beta_{n-1} \}} \left(\prod^{n-1}_{k=0} J_{\alpha_{k} \beta_{k}} \right)\la \alpha_{0} (w_{0}) \cdots \alpha_{n-1} (w_{n-1}) \ra \la \beta_{0} (w'_{0}) \cdots \beta_{n-1} (w'_{n-1}) \ra, \no \label{eq:main}
\eea

where
\be
J_{\alpha_{k} \beta_{k}}  =\sum_{a_{k}} p_{a_{k}} \left( \f{ \la X_{a_{k}}(w_{k}) X_{a_{k}}(\hat{w}_{k}) \ra_{\Sigma_{n}}}{ \la X_{a_{k}}(w_{0}) X_{a_{k}}(\hat{w}_{0}) \ra_{\Sigma_{1}}}\right)\left(  \f{ \la Y_{a_{k}}(w'_{k}) Y_{a_{k}}(\hat{w}'_{k}) \ra_{\Sigma_{n}}}{ \la Y_{a_{k}}(w'_{0}) Y_{a_{k}}(\hat{w}'_{0}) \ra_{\Sigma_{1}}} \right)  C_{X_{a_{k}}X_{a_{k}}   \alpha_{k}}C_{Y_{a_{k}}Y_{a_{k}} \beta_{k}} (2l)^{\Delta_{\alpha_{k}}+\Delta_{\beta_{k}}}. \label{eq:Jab}
\ee

When the subsystem size $l$ is small, $\alpha_{k}$ can only be either identity $1$ or the first non trivial primary $O$ with the scaling dimension $\Delta$,
$\alpha_{k} \in \{1, O\}$, and similarly, $\beta_{k} \in \{1, O\}$.  This implies that we have the following expansion of $ {\rm tr} \;  \sigma_{AB}^n$ (\ref{eq:main}) in terms of $l^{\Delta}$,
\be
{\rm tr} \; \sigma_{AB}^n\cdot \f{(Z^{(1)}_{A} Z^{(1)}_{B})^n}{Z^{(n)}_{A}\; Z^{(n)}_{B}}= L^{(n)}_{0} + L^{(n)}_{2} \; ( l)^{2\Delta} + L^{(n)}_{3} \; ( l)^{3\Delta}+ L^{(n)}_{4}\; ( l)^{4\Delta}  +\cdots
\ee

In the next few subsections we calculate these coefficients.
\subsection{$L^{(n)}_{0}$: the first law part}

Only the trivial operator configuration can contribute to the coefficient
\be
\{  \alpha_{0}, \cdots \alpha_{n-1} \} =\{ 1, \cdots 1 \}, \quad \{  \beta_{0}, \cdots \beta_{n-1} \} =\{ 1, \cdots 1 \}
\ee

therefore  $L^{(n)}_{0} =J_{11}^n$, and
\begin{align}
\f{\p }{\p n}L^{(n)}_{0} \big|_{n=1}&= \sum_{a} p_{a} \left[ \f{\p }{\p n}\left(\f{ \la X_{a}(w_{0}) X_{a}(\hat{w}_{0}) \ra_{\Sigma_{n}}}{ \la X_{a}(w_{0}) X_{a}(\hat{w}_{0}) \ra_{\Sigma_{1}}} \right) +\f{\p }{\p n}\left(\f{ \la Y_{a}(w'_{0}) Y_{a}(\hat{w}'_{0}) \ra_{\Sigma_{n}}}{ \la Y_{a}(w'_{0}) Y_{a}(\hat{w}'_{0}) \ra_{\Sigma_{1}}} \right)\right] \big|_{n=1}
\nonumber \\[+10pt]
&= -\sum_{a} p_{a} \left( \la K^{0}_{A} (\rho^{a}_{A}-\rho^{0}_{A} )\ra+ \la K^{0}_{B} (\rho^{a}_{B}-\rho^{0}_{B} ) \ra \right), \label{eq:fir}
\end{align}

where $K^{0}_{A}, K^{0}_{B}$ is vacuum modular Hamiltonian of  region A and B respectively. This part  is just an analog of the first law part of   excited state entanglement entropy.

\subsection{$L^{(n)}_{1}$}

Configurations in which only one  non trivial operator is present are not allowed because every  vacuum one point function vanishes.   Therefore $L^{(n)}_{1} =0 $.

\subsection{$L^{(n)}_{2}$}

In this case two types of operator configuration can contribute to the coefficient. One is
\be
\{  \alpha_{0}, \cdots \alpha_{n-1} \} =\{ 1, \cdots  O_{q_{1}}, \cdots O_{j},   \cdots 1 \}, \quad \{  \beta_{0}, \cdots \beta_{n-1} \} =\{ 1, \cdots 1 \}, \; q_{1} <j
\ee

and

\be
\{  \alpha_{0}, \cdots \alpha_{n-1} \} = \{ 1, \cdots 1 \}, \quad \{  \beta_{0}, \cdots \beta_{n-1} \} =\{ 1, \cdots  O_{q_{2}}, \cdots O_{k},   \cdots 1 \}, \; q_{2} <k
\ee

In both cases  there are  two  non trivial operators.
\be
L^{(n)}_{2} (l)^{2\Delta} =\f{J_{11}^{n-2}}{2}\sum_{q_{1}=0}^{n-1} \sum_{j =0 \neq q_{1}}^{n-1} J_{O_{q_{1}} 1} J_{O_{j} 1} \la O(w_{q_{1}})O(w_{j}) \ra + \f{J_{11}^{n-2}}{2}\sum_{q_{2}=0}^{n-1} \sum_{k =0 \neq q_{2}}^{n-1} J_{1O_{q_{2}} } J_{1O_{k}} \la O(w_{q_{2}})O(w_{k}) \ra.
\ee

We are only  interested in $n \rightarrow 1$ limit. In this case we can set $n=1$ in $J_{O1}$ as the sum of two point function  $ \sum_{j}\la O(w_{q_{1}})O(w_{j}) \ra$ is already proportional to $n-1$ \cite{Calabrese:2009ez},

\be
f(\Delta,n)= \sum_{j=1}^{n-1} \la O(w_{0})O(w_{j}) \ra= \sum_{k=1}^{n-1} \f{1}{\left(2n\sin \f{\pi k}{n}  \right)^{2\Delta}} \rightarrow (n-1) \f{\Gamma(3/2) \Gamma(\Delta+1)}{2^{2\Delta}\Gamma( \Delta +3/2)}, \quad n \rightarrow 1,
\ee
therefore
\be
J_{O1}  = (2l)^{\Delta} \sum_{a} p_{a} C_{X_{a}X_{a}O} , \quad J_{1O} = (2l)^{\Delta}\sum_{a} p_{a} C_{Y_{a}Y_{a}O}, \quad J_{11}=1.
\ee

Combining them, we conclude,
\be
(l)^{2\Delta}\f{\p }{\p n}L^{(n)}_{2} \big|_{n=1} = \f{\Gamma(3/2) \Gamma(\Delta+1)}{2\Gamma( \Delta +3/2)} \left[\left(\sum_{a} p_{a} C_{X_{a}X_{a}O} \right)^{2} + \left(\sum_{a} p_{a} C_{Y_{a}Y_{a}O} \right)^{2} \right] (l)^{2\Delta}. \label{eq:quad}
\ee

\subsection{$L^{(n)}_{3}$}

In this term again we have two types of contributions
\be
\{  \alpha_{0}, \cdots \alpha_{n-1} \} =\{ 1, \cdots  O_{q_{1}}, \cdots O_{q_{2}},   \cdots O_{q_{3}} \cdots 1 \}, \quad \{  \beta_{0}, \cdots \beta_{n-1} \} =\{ 1, \cdots 1 \}, \; q_{1} <q_{2}<q_{3}
\ee

and
\be
\{  \alpha_{0}, \cdots \alpha_{n-1} \} =\{ 1, \cdots 1 \} , \quad \{  \beta_{0}, \cdots \beta_{n-1} \} =\{ 1, \cdots  O_{p_{1}}, \cdots O_{p_{2}},   \cdots O_{p_{3}} \cdots 1 \}, \; p_{1} <p_{2}<p_{3}.
\ee

As in the case of $L^{(n)}_{2}$, the first contribution generates the  cubic order  of  the von Neuman entrpy on region A, $S(\sigma_{A})$  which was explained in  (\ref{eq: eeformula}) , and similarly the second contribution generates the cubic order of $S(\sigma_{B})$. Therefore we conclude,

\be
l^{3\Delta} \f{\p }{\p n}L^{(n)}_{3} \big|_{n=1}=C_{OOO}b_{\Delta}l^{3\Delta} \left[\left(\sum_{a} p_{a}C_{X_{a}X_{a}O} \right)^3+\left(\sum_{a} p_{a} C_{Y_{a}Y_{a}O} \right)^3 \right]. \nonumber \\
\ee

\subsection{$L^{(n)}_{4}$}

In this case we have
\be
\{  \alpha_{0}, \cdots \alpha_{n-1} \} =\{ 1, \cdots  O_{q_{1}}, \cdots O_{j},   \cdots 1 \}, \quad \{  \beta_{0}, \cdots \beta_{n-1} \} =\{ 1, \cdots  O_{q_{2}}, \cdots O_{k},   \cdots 1 \}
\ee

and
\be
(l)^{4\Delta} L^{(n)}_{4} = \f{1}{4} \sum_{q_{1}=0}^{n-1} \sum_{j =0 \neq q_{1}}^{n-1}   \sum_{q_{2}=0}^{n-1} \sum_{k =0 \neq q_{2}}^{n-1} I_{q_{1}, q_{2}}^{j,k}. \label{eq:quadsum}
\ee

The precise form of $ I_{q_{1}, q_{2}}^{j,k}$  highly depends on   the value of the indices.   For example, when
$(j=q_{2}, k=q_{1})$,
\be
I^{q_{2},q_{1}}_{q_{1},q_{2}} = J_{OO}^2 C(q_{1}-q_{2})^2,   \quad C(q_{1}-q_{2})\equiv \la O(w_{q_{1}}) O(w_{q_{2}}) \ra
\ee

with
\be
J_{OO} = (2l)^{2\Delta} \sum_{a} p_{a} C_{X_{a}X_{a}O}C_{Y_{a}Y_{a}O}.
\ee

We can compare this expression to (51) of \cite{Ugajin:2016opf} . They can be identified by the replacement
 $ \la \mathcal{O}_{\alpha}  \mathcal{O}_{\beta} \ra \rightarrow J_{OO}^2$.

When $\{q_{1} \neq q_{2} \neq j \neq k \}.$

\be
I^{j,k}_{q_{1},q_{2}}= J_{O1}^2 J_{1O}^2 C(q_{1} -j) C(q_{2} -k)
\ee
Again this can be compare to (59) of\cite{Ugajin:2016opf}, and they are identified by $\la\mathcal{O}_{\alpha} \ra \la \mathcal{O}_{\beta} \ra \rightarrow J_{O1} J_{1O}$.

The strategy to calculate the
 sum (\ref{eq:quadsum})  is almost same as the calculation of Appendix A of \cite{Ugajin:2016opf}  ie,   first computing the sum with respect to $j,k$ with fixed $q_{1}, q_{2}$,
\be
I_{q_{1}, q_{2}}  = \sum_{j =0 \neq q_{1}}^{n-1} \sum_{k =0 \neq q_{2}}^{n-1}    I_{q_{1}, q_{2}}^{j,k},
\ee

then performing the sum with respect to  $q_{1}, q_{2}$.

Indeed, we can easily convince ourself that we can derive the result of the sum (\ref{eq:quadsum}) from (69) of Appendix A of \cite{Ugajin:2016opf} , just by the replacing $ \la \mathcal{O}_{\alpha}  \mathcal{O}_{\beta} \ra$ in \cite{Ugajin:2016opf} to $J_{OO}$ and  $\la\mathcal{O}_{\alpha} \ra \la \mathcal{O}_{\beta} \ra$  to
$J_{O1} J_{1O}$. Therefore the final result is
\bea
&& (l)^{4\Delta} \f{\p}{\p n}L^{(n)}_{4}  \no
&&=  \f{\Gamma(3/2) \Gamma(2\Delta+1)}{2^{4\Delta+1}\Gamma( 2\Delta +3/2)} \left( J_{OO} -J_{O1} J_{O1}\right)^2 \nonumber \\
&&=  \f{\Gamma(3/2) \Gamma(2\Delta+1)}{2\Gamma( 2\Delta +3/2)} \left[ \sum_{a} p_{a} C_{X_{a}X_{a}O}C_{Y_{a}Y_{a}O}-\left(\sum_{a} p_{a} C_{X_{a}X_{a}O} \right)\left(\sum_{a} p_{a} C_{Y_{a}Y_{a}O} \right)\right]^{2}  (l)^{4\Delta}. \no
 \label{eq: fouth}
\eea

\subsection{$L^{(n)}_{5}$}

We similarly have $L^{(n)}_{5}$ term. This term can be relevant in section \ref{sec;nextleading} in which we compute the relative entropy up to $l^{6\Delta}$ term by assuming the locally vacuum condition. However  if we assume this condition,  $L^{(n)}_{5}$ term is vanishing, therefore we can ignore this term.

\subsection{$L^{(n)}_{6}$}

We can also compute the  one more higher term $L^{n}_{6}$ once we assume the locally vacuum condition \ref{eq;locvac1}.

From the OPE expansion (\ref{eq:main}) and the condition \ref{eq;locvac1},  the result is,

\be
(l)^{6\Delta} L^{(n)}_{6}= J_{OO}^3  \left[ \f{1}{6} \sum_{\{q_{1}, q_{2}, q_{3}\}}  \la O_{q_{1}}O_{q_{2}}O_{q_{3}}. \ra^{2}_{\Sigma_{n}} \right].  \label{eq: 6th}
\ee

It is hard to directly perform the sum in right hand side and analytically continue the result in $n$. However we can read off the outcome from (5.15) of \cite{Sarosi:2017rsq} where they computed  the entangle entropy of an excited state at cubic order,
\be
\lim_{n \rightarrow 1} \f{1}{n-1} \sum_{\{q_{1}, q_{2}, q_{3}\}}  \la O_{q_{1}}O_{q_{2}}O_{q_{3}} \ra_{\Sigma_{n}}  = -C_{OOO} \f{\Gamma(\f{1+\Delta}{2})^3}{12\pi \Gamma(\f{3+3\Delta}{2})}.
\ee

In our case (\ref{eq: 6th}) we have
\be
\lim_{n \rightarrow 1} \f{1}{n-1} \sum_{\{q_{1}, q_{2}, q_{3}\}}  \la O_{q_{1}}O_{q_{2}}O_{q_{3}} \ra^{2}_{\Sigma_{n}}  = -C_{OOO}^2 \f{\Gamma(\f{1+2\Delta}{2})^3}{12\pi \Gamma(\f{3+6\Delta}{2})}.
\ee

Therefore
\be
-S ({\displaystyle \sum_{a} } \;p_{a}\;  \rho^{a}_{A} \otimes \rho^{a}_{B}) \Big|_{l^{6\Delta}}= - (l)^{6\Delta} \left(\sum_{a} p_{a} \la \rho^{a} O \ra^2 \right)^3  C_{OOO}^2 \f{\Gamma(\f{1+2\Delta}{2})^3}{12\pi \Gamma(\f{3+6\Delta}{2})}.
\ee

By defining
\be
Z\equiv (l)^{2\Delta}\sum_{a} p_{a} \la \rho^{a} O \ra^2 , \quad  d_{\Delta} \equiv 2^{6\Delta} \f{\Gamma(\f{1+2\Delta}{2})^3}{12\pi \Gamma(\f{3+6\Delta}{2})}
\ee

we write
\be
-S ({\displaystyle \sum_{a} } \;p_{a}\;  \rho^{a}_{A} \otimes \rho^{a}_{B}) \Big|_{l^{6\Delta}} = -\left( d_{\Delta} C_{OOO}^2 \right) Z^{3} \label{eq:aaa}.
\ee

\subsection{The final result}
By plugging (\ref{eq:fir}),  (\ref{eq:quad}),  (\ref{eq: fouth}) we obtain the expression of the von Neumann entropy up to $l^{4\Delta}$ order,
\begin{flalign}
-S( \sum_{a}\;  p_{a} \; \rho^{a}_{A}  \rho^{a}_{B})&=  \f{\p}{\p n} \left[\left(L^{(n)}_{0} + L^{(n)}_{2} \; ( l)^{2\Delta} +
L^{(n)}_{3} \; ( l)^{3\Delta} + L^{(n)}_{4}\; ( l)^{4\Delta}  +\cdots \right)Z^{(n)}_{A}\; Z^{(n)}_{B}\right]\big|_{n=1} \nonumber \\
&= -\sum_{a} p_{a} \left( \la K^{0}_{A} \rho^{a}_{A} \ra+ \la K^{0}_{B} \rho^{a}_{B} \ra \right) \nonumber \\
&+(l)^{2\Delta}a_{\Delta} \left[\left(\sum_{a} p_{a} C_{X_{a}X_{a}O} \right)^{2} +\left(\sum_{a} p_{a} C_{Y_{a}Y_{a}O} \right)^{2} \right]  \nonumber \\
&-C_{OOO}b_{\Delta}l^{3\Delta} \left[\left(\sum_{a} p_{a} C_{X_{a}X_{a}O}   \right)^3+\left(\sum_{a} p_{a} \la  C_{Y_{a}Y_{a}O}  \ra \right)^3 \right] \nonumber \\
&+ (l)^{4\Delta}a_{2\Delta} \left[ \sum_{a} p_{a} C_{X_{a}X_{a}O}C_{Y_{a}Y_{a}O}-\left(\sum_{a} p_{a} C_{X_{a}X_{a}O} \right)\left(\sum_{a} p_{a} C_{Y_{a}Y_{a}O} \right)\right]^{2}.
\end{flalign}

We can see that up to the order of $l^{2\Delta}$ the entropy splits, $S= S(\sum p_{a} \rho^{a}_{A})+ S(\sum p_{a} \rho^{a}_{B})$.  However this no longer holds at the $l^{4\Delta}$ order.

It can also be written in terms of the reduced density matrices $\{ \rho^{a}_{A}, \rho^{a}_{B} \}$.
\begin{align}
-S( \sum_{a}\;  p_{a} \; \rho^{a}_{A}  \rho^{a}_{B})&=
 -\sum_{a} p_{a} \left( \la K^{0}_{A} \rho^{a}_{A} \ra+ \la K^{0}_{B} \rho^{a}_{B} \ra \right) \nonumber \\
&+a_{\Delta}\left(l\right)^{2\Delta} \left[\left(\sum_{a} p_{a} \la \rho^{a}_{A}O \ra\right)^2+\left(\sum_{a} p_{a} \la \rho^{a}_{B}O \ra \right)^2\right] \nonumber \\
&-C_{OOO}b_{\Delta}l^{3\Delta} \left[\left(\sum_{a} p_{a} \la \rho^{a}_{A}O \ra\right)^3+\left(\sum_{a} p_{a} \la \rho^{a}_{B}O \ra \right)^3 \right] \nonumber \\
&+ a_{2\Delta}\left(l\right)^{4\Delta}\left[ \sum_{a}p_{a}  \la \rho^{a}_{A} O_{A} \ra \la \rho^{a}_{B} O_{B}\ra -  \left(\sum_{a} p_{a} \la \rho^{a}_{A}O \ra\right) \left(\sum_{a} p_{a} \la \rho^{a}_{B}O \ra \right)\right]^2.
\end{align}

The second term is

\begin{align}
{\rm tr} \left[ \sum_{a} p_{a}\; \rho^{a}_{A} \; \rho^{i}_{B} K^{0}_{AB} \right]& =\sum_{i} p_{i} \left[ \la K^{0}_{A} \ra_{i} +\la K^{0}_{B} \ra_{i} \right]  \nonumber \\
&-  2 a_{2\Delta} (l)^{2\Delta} \left(\f{l}{R}\right)^{2\Delta} \sum_{a} p_{a} \left[ \la \rho^{a}_{A} O_{A} \ra  \la \rho^{a}_{B}  O_{B} \ra  \right] + I_{AB}.  \\
\end{align}

The net result is
\begin{align}
S(\sigma_{AB}||\rho_{AB})&=a_{\Delta}\left(l\right)^{2\Delta} \left[\left(\sum_{a} p_{a} \la \rho^{a}_{A}O \ra\right)^2+\left(\sum_{a} p_{a} \la \rho^{a}_{B}O \ra \right)^2\right] \nonumber \\
&-C_{OOO}b_{\Delta}l^{3\Delta} \left[\left(\sum_{a} p_{a} \la \rho^{a}_{A}O \ra\right)^3+\left(\sum_{a} p_{a} \la \rho^{a}_{B}O \ra \right)^3 \right] \nonumber \\
&+ a_{2\Delta}\left(l\right)^{4\Delta}\left[ \sum_{a}p_{a}  \la \rho^{a}_{A} O_{A} \ra \la \rho^{a}_{B} O_{B}\ra -  \left(\sum_{a} p_{a} \la \rho^{a}_{A}O \ra\right) \left(\sum_{a} p_{a} \la \rho^{a}_{B}O \ra \right)\right]^2 \nonumber \\
&-  2 a_{2\Delta} (l)^{2\Delta} \left(\f{l}{R}\right)^{2\Delta} \sum_{a} p_{a} \left[ \la \rho^{a}_{A} O_{A} \ra  \la \rho^{a}_{B}  O_{B} \ra  \right] + I_{AB}.
\end{align}

\section{On a replacement rule}\label{sec;replacement}

In the body  of the paper, we used the fact that $S(\rho_{AB}) $ is related to  $S(\sigma_{AB})$  by the replacement,
\be
W(\rho_{AB}) ={\rm tr} \left[ \rho_{AB} O_{A} O_{B} \right] \rightarrow \left[ \sigma_{AB} O_{A} O_{B} \right] =Z(\sigma_{AB}).
\ee

In this appendix we prove this prescription. For simplicity we consider the case where $\rho_{AB}$  is the reduced density matrix of a pure state,
\be
\rho_{AB} ={\rm tr}_{(AB)^{c}} |V \ra \la V|,
\ee

and for $\sigma_{AB} $, (\ref{eq:two density}).

The R\'enyi entropy ${\rm tr} \rho_{AB}^{n}$ has an expression in terms of a correlation function of
 the twist defect $D_{n} $ \cite{Cardy:2013nua},
\be
{\rm tr} \rho_{AB}^{n} = \la V(\infty)^{\otimes n} D_{n} (A) D_{n}(B) V(0)^{\otimes n} \ra, \label{eq;renyicorr}
\ee
the correlation function is evaluated on the cyclic orbifold $(CFT)^{\otimes n} /Z_{n} $ of the original CFT.
Here we take  $\la V(\infty) V(0) \ra =1$.
In the small subsystem size limit $|A|, |B| \rightarrow 0$ one can expand the twist defect in terms of local operators,
\be
D_{n} (A)= \sum_{\{O_{k}\}}\;  l^{\sum_{k=0}^{n-1} \Delta_{k} }\; \la \prod^{n-1}_{k=0} O_{k}(A) \ra_{\Sigma_{n}} \; \prod^{n-1}_{k=0} O_{k}(A),  \label{eq;locexpansion}
\ee
here $\la \cdots \ra_{\Sigma_{n}}$ indicates that we evaluate the correlation function on the branched space $\Sigma_{n}$, with a cut on the region $A$. By plugging this expansion (\ref{eq;locexpansion}) into (\ref{eq;renyicorr}), we get
\be
{\rm tr} \rho_{AB}^{n} = \sum_{\{O^{A}_{k}, \tilde{O}^{B}_{k}\}} \;  l^{\sum_{k=0}^{n-1} (\Delta_{k}+\tilde{\Delta}_{k}) }\;
\la \prod^{n-1}_{k=0} O_{k}(A) \ra_{\Sigma_{n}} \la \prod^{n-1}_{k=0} \tilde{O}_{k}(B) \ra_{\Sigma_{n}} \prod^{n-1}_{k=0} \la V(\infty) O^{A}_{k} \tilde{O}^{B}_{k} V(0) \ra, \label{eq;rhoren}
\ee
notice  in general $O^{A}_{k} \neq \tilde{O}^{B}_{k}$.
On the other hand  from (\ref{eq:main}),
\be
{\rm tr} \;\sigma_{AB}^{n} = \sum_{\{O^{A}_{k}, \tilde{O}^{B}_{k}\}}
\la \prod^{n-1}_{k=0} O_{k}(A) \ra_{\Sigma_{n}} \la \prod^{n-1}_{k=0} \tilde{O}_{k}(B) \ra_{\Sigma_{n}}\prod^{n-1}_{k=0} J_{O_{k} \tilde{O}_{k}},\label{eq;sigmaren}
\ee

with (\ref{eq:Jab})

\be
J_{O_{k} \tilde{O}_{k}}  =\sum_{a_{k}} p_{a_{k}} \left( \f{ \la X_{a_{k}}(w_{k}) X_{a_{k}}(\hat{w}_{k}) \ra_{\Sigma_{n}}}{ \la X_{a_{k}}(w_{0}) X_{a_{k}}(\hat{w}_{0}) \ra_{\Sigma_{1}}}\right)\left(  \f{ \la Y_{a_{k}}(w'_{k}) Y_{a_{k}}(\hat{w}'_{k}) \ra_{\Sigma_{n}}}{ \la Y_{a_{k}}(w'_{0}) Y_{a_{k}}(\hat{w}'_{0}) \ra_{\Sigma_{1}}} \right)  C_{X_{a_{k}}X_{a_{k}}   O_{k}}C_{Y_{a_{k}}Y_{a_{k}} \tilde{O}_{k}} (2l)^{\Delta_{O_{k}}+\Delta_{\tilde{O}_{k}}}. \label{eq:Jabc}
\ee

In the $n \rightarrow 1$ limit, these two expressions (\ref{eq;rhoren}),(\ref{eq;sigmaren}) are related by the identification,
\begin{align}
{\rm tr}\left[ \rho_{AB} O_{k}(A) \tilde{O}_{k}(B) \right] = \la V(\infty) O^{A}_{k} \tilde{O}^{B}_{k} V(0) \ra
\leftrightarrow  \sum_{a_{k}}p_{a_{k}} C_{X_{a_{k}}X_{a_{k}} O^{A}_{k}} C_{Y_{a_{k}}Y_{a_{k}} \tilde{O}^{B}_{k}}=
{\rm tr}\left[ \sigma_{AB} O_{k} \tilde{O}_{k} \right].
\end{align}

\bibliographystyle{utphys}
\bibliography{separable}

\providecommand{\href}[2]{#2}\begingroup\raggedright\begin{thebibliography}{10}

\bibitem{DonaldHorodeckiRudolph01:TheUniqueness}
M.~J. Donald, M.~Horodecki, and O.~Rudolph, ``The uniqueness theorem for
  entanglement measures,'' \href{http://dx.doi.org/10.1023/A:1026654312961,
  10.4310/ATMP.1998.v2.n2.a1}{{\em Journal of Mathematical Physics} {\bfseries
  43} no.~9, (2002) 4252--4272},
  \href{http://arxiv.org/abs/quant-ph/0105017}{{\ttfamily
  arXiv:quant-ph/0105017 [quant-ph]}}.

\bibitem{Calabrese:2009qy}
P.~Calabrese and J.~Cardy, ``{Entanglement entropy and conformal field
  theory},'' \href{http://dx.doi.org/10.1088/1751-8113/42/50/504005}{{\em J.
  Phys.} {\bfseries A42} (2009) 504005},
\href{http://arxiv.org/abs/0905.4013}{{\ttfamily arXiv:0905.4013
  [cond-mat.stat-mech]}}.
%%CITATION = ARXIV:0905.4013;%%.

\bibitem{Casini:2009sr}
H.~Casini and M.~Huerta, ``{Entanglement entropy in free quantum field
  theory},'' \href{http://dx.doi.org/10.1088/1751-8113/42/50/504007}{{\em J.
  Phys.} {\bfseries A42} (2009) 504007},
\href{http://arxiv.org/abs/0905.2562}{{\ttfamily arXiv:0905.2562 [hep-th]}}.
%%CITATION = ARXIV:0905.2562;%%.

\bibitem{Nishioka:2009un}
T.~Nishioka, S.~Ryu, and T.~Takayanagi, ``{Holographic Entanglement Entropy: An
  Overview},'' \href{http://dx.doi.org/10.1088/1751-8113/42/50/504008}{{\em J.
  Phys.} {\bfseries A42} (2009) 504008},
\href{http://arxiv.org/abs/0905.0932}{{\ttfamily arXiv:0905.0932 [hep-th]}}.
%%CITATION = ARXIV:0905.0932;%%.

\bibitem{Rangamani:2016dms}
M.~Rangamani and T.~Takayanagi, ``{Holographic Entanglement Entropy},''
  \href{http://dx.doi.org/10.1007/978-3-319-52573-0}{{\em Lect. Notes Phys.}
  {\bfseries 931} (2017) pp.1--246},
\href{http://arxiv.org/abs/1609.01287}{{\ttfamily arXiv:1609.01287 [hep-th]}}.
%%CITATION = ARXIV:1609.01287;%%.

\bibitem{Nishioka:2018khk}
T.~Nishioka, ``{Entanglement entropy: holography and renormalization group},''
\href{http://arxiv.org/abs/1801.10352}{{\ttfamily arXiv:1801.10352 [hep-th]}}.
%%CITATION = ARXIV:1801.10352;%%.

\bibitem{Horodecki:2009zz}
R.~Horodecki, P.~Horodecki, M.~Horodecki, and K.~Horodecki, ``{Quantum
  entanglement},'' \href{http://dx.doi.org/10.1103/RevModPhys.81.865}{{\em Rev.
  Mod. Phys.} {\bfseries 81} (2009) 865--942},
\href{http://arxiv.org/abs/quant-ph/0702225}{{\ttfamily arXiv:quant-ph/0702225
  [quant-ph]}}.
%%CITATION = QUANT-PH/0702225;%%.

\bibitem{GOQ}
I.~Bengtsson and K.~Zyczkowski, {\em Geometry of Quantum States}.
\newblock Cambridge university press, 2006.

\bibitem{2002JMP....43.4286T}
B.~M. {Terhal}, M.~{Horodecki}, D.~W. {Leung}, and D.~P. {DiVincenzo}, ``{The
  entanglement of purification},''
  \href{http://dx.doi.org/10.1063/1.1498001}{{\em Journal of Mathematical
  Physics} {\bfseries 43} (Sept., 2002) 4286--4298},
  \href{http://arxiv.org/abs/quant-ph/0202044}{{\ttfamily quant-ph/0202044}}.

\bibitem{Takayanagi:2017knl}
T.~Takayanagi and K.~Umemoto, ``{Holographic Entanglement of Purification},''
  \href{http://dx.doi.org/10.1038/s41567-018-0075-2}{{\em Nature Phys.}
  {\bfseries 14} no.~6, (2018) 573--577},
\href{http://arxiv.org/abs/1708.09393}{{\ttfamily arXiv:1708.09393 [hep-th]}}.
%%CITATION = ARXIV:1708.09393;%%.

\bibitem{Nguyen:2017yqw}
P.~Nguyen, T.~Devakul, M.~G. Halbasch, M.~P. Zaletel, and B.~Swingle,
  ``{Entanglement of purification: from spin chains to holography},''
  \href{http://dx.doi.org/10.1007/JHEP01(2018)098}{{\em JHEP} {\bfseries 01}
  (2018) 098},
\href{http://arxiv.org/abs/1709.07424}{{\ttfamily arXiv:1709.07424 [hep-th]}}.
%%CITATION = ARXIV:1709.07424;%%.

\bibitem{Bhattacharyya:2018sbw}
A.~Bhattacharyya, T.~Takayanagi, and K.~Umemoto, ``{Entanglement of
  Purification in Free Scalar Field Theories},''
  \href{http://dx.doi.org/10.1007/JHEP04(2018)132}{{\em JHEP} {\bfseries 04}
  (2018) 132},
\href{http://arxiv.org/abs/1802.09545}{{\ttfamily arXiv:1802.09545 [hep-th]}}.
%%CITATION = ARXIV:1802.09545;%%.

\bibitem{Bao:2017nhh}
N.~Bao and I.~F. Halpern, ``{Holographic Inequalities and Entanglement of
  Purification},'' \href{http://dx.doi.org/10.1007/JHEP03(2018)006}{{\em JHEP}
  {\bfseries 03} (2018) 006},
\href{http://arxiv.org/abs/1710.07643}{{\ttfamily arXiv:1710.07643 [hep-th]}}.
%%CITATION = ARXIV:1710.07643;%%.

\bibitem{Blanco:2018riw}
D.~Blanco, M.~Leston, and G.~Pérez-Nadal, ``{Gravity from entanglement for
  boundary subregions},'' \href{http://dx.doi.org/10.1007/JHEP06(2018)130}{{\em
  JHEP} {\bfseries 2018} (2018) 130},
\href{http://arxiv.org/abs/1803.01874}{{\ttfamily arXiv:1803.01874 [hep-th]}}.
%%CITATION = ARXIV:1803.01874;%%.

\bibitem{Hirai:2018jwy}
H.~Hirai, K.~Tamaoka, and T.~Yokoya, ``{Towards Entanglement of Purification
  for Conformal Field Theories},''
  \href{http://dx.doi.org/10.1093/ptep/pty063}{{\em PTEP} {\bfseries 2018}
  no.~6, (2018) 063B03},
\href{http://arxiv.org/abs/1803.10539}{{\ttfamily arXiv:1803.10539 [hep-th]}}.
%%CITATION = ARXIV:1803.10539;%%.

\bibitem{Espindola:2018ozt}
R.~Espindola, A.~Guijosa, and J.~F. Pedraza, ``{Entanglement Wedge
  Reconstruction and Entanglement of Purification},''
\href{http://arxiv.org/abs/1804.05855}{{\ttfamily arXiv:1804.05855 [hep-th]}}.
%%CITATION = ARXIV:1804.05855;%%.

\bibitem{Bao:2018gck}
N.~Bao and I.~F. Halpern, ``{Conditional and Multipartite Entanglements of
  Purification and Holography},''
\href{http://arxiv.org/abs/1805.00476}{{\ttfamily arXiv:1805.00476 [hep-th]}}.
%%CITATION = ARXIV:1805.00476;%%.

\bibitem{Nomura:2018kji}
Y.~Nomura, P.~Rath, and N.~Salzetta, ``{Pulling the Boundary into the Bulk},''
  \href{http://dx.doi.org/10.1103/PhysRevD.98.026010}{{\em Phys. Rev.}
  {\bfseries D98} no.~2, (2018) 026010},
\href{http://arxiv.org/abs/1805.00523}{{\ttfamily arXiv:1805.00523 [hep-th]}}.
%%CITATION = ARXIV:1805.00523;%%.

\bibitem{Umemoto:2018jpc}
K.~Umemoto and Y.~Zhou, ``{Entanglement of Purification for Multipartite States
  and its Holographic Dual},''
\href{http://arxiv.org/abs/1805.02625}{{\ttfamily arXiv:1805.02625 [hep-th]}}.
%%CITATION = ARXIV:1805.02625;%%.

\bibitem{2008PhRvL.101k0501V}
G.~{Vidal}, ``{Class of Quantum Many-Body States That Can Be Efficiently
  Simulated},'' \href{http://dx.doi.org/10.1103/PhysRevLett.101.110501}{{\em
  Physical Review Letters} {\bfseries 101} no.~11, (Sept., 2008) 110501},
  \href{http://arxiv.org/abs/quant-ph/0610099}{{\ttfamily quant-ph/0610099}}.

\bibitem{Calabrese:2012ew}
P.~Calabrese, J.~Cardy, and E.~Tonni, ``{Entanglement negativity in quantum
  field theory},'' \href{http://dx.doi.org/10.1103/PhysRevLett.109.130502}{{\em
  Phys. Rev. Lett.} {\bfseries 109} (2012) 130502},
\href{http://arxiv.org/abs/1206.3092}{{\ttfamily arXiv:1206.3092
  [cond-mat.stat-mech]}}.
%%CITATION = ARXIV:1206.3092;%%.

\bibitem{Calabrese:2012nk}
P.~Calabrese, J.~Cardy, and E.~Tonni, ``{Entanglement negativity in extended
  systems: A field theoretical approach},''
  \href{http://dx.doi.org/10.1088/1742-5468/2013/02/P02008}{{\em J. Stat.
  Mech.} {\bfseries 1302} (2013) P02008},
\href{http://arxiv.org/abs/1210.5359}{{\ttfamily arXiv:1210.5359
  [cond-mat.stat-mech]}}.
%%CITATION = ARXIV:1210.5359;%%.

\bibitem{Calabrese:2013mi}
P.~Calabrese, L.~Tagliacozzo, and E.~Tonni, ``{Entanglement negativity in the
  critical Ising chain},''
  \href{http://dx.doi.org/10.1088/1742-5468/2013/05/P05002}{{\em J. Stat.
  Mech.} {\bfseries 1305} (2013) P05002},
\href{http://arxiv.org/abs/1302.1113}{{\ttfamily arXiv:1302.1113
  [cond-mat.stat-mech]}}.
%%CITATION = ARXIV:1302.1113;%%.

\bibitem{Vedral:1997qn}
V.~Vedral, M.~B. Plenio, M.~A. Rippin, and P.~L. Knight, ``{Quantifying
  entanglement},'' \href{http://dx.doi.org/10.1103/PhysRevLett.78.2275}{{\em
  Phys. Rev. Lett.} {\bfseries 78} (1997) 2275--2279},
\href{http://arxiv.org/abs/quant-ph/9702027}{{\ttfamily arXiv:quant-ph/9702027
  [quant-ph]}}.
%%CITATION = QUANT-PH/9702027;%%.

\bibitem{Vedral:1997hd}
V.~Vedral and M.~B. Plenio, ``{Entanglement measures and purification
  procedures},'' \href{http://dx.doi.org/10.1103/PhysRevA.57.1619}{{\em Phys.
  Rev.} {\bfseries A57} (1998) 1619--1633},
\href{http://arxiv.org/abs/quant-ph/9707035}{{\ttfamily arXiv:quant-ph/9707035
  [quant-ph]}}.
%%CITATION = QUANT-PH/9707035;%%.

\bibitem{Sarosi:2016oks}
G.~S\'arosi and T.~Ugajin, ``{Relative entropy of excited states in two
  dimensional conformal field theories},''
  \href{http://dx.doi.org/10.1007/JHEP07(2016)114}{{\em JHEP} {\bfseries 07}
  (2016) 114},
\href{http://arxiv.org/abs/1603.03057}{{\ttfamily arXiv:1603.03057 [hep-th]}}.
%%CITATION = ARXIV:1603.03057;%%.

\bibitem{Lashkari:2016idm}
N.~Lashkari, J.~Lin, H.~Ooguri, B.~Stoica, and M.~Van~Raamsdonk,
  ``{Gravitational positive energy theorems from information inequalities},''
  \href{http://dx.doi.org/10.1093/ptep/ptw139}{{\em PTEP} {\bfseries 2016}
  no.~12, (2016) 12C109},
\href{http://arxiv.org/abs/1605.01075}{{\ttfamily arXiv:1605.01075 [hep-th]}}.
%%CITATION = ARXIV:1605.01075;%%.

\bibitem{Sarosi:2016atx}
G.~S\'arosi and T.~Ugajin, ``{Relative entropy of excited states in conformal
  field theories of arbitrary dimensions},''
  \href{http://dx.doi.org/10.1007/JHEP02(2017)060}{{\em JHEP} {\bfseries 02}
  (2017) 060},
\href{http://arxiv.org/abs/1611.02959}{{\ttfamily arXiv:1611.02959 [hep-th]}}.
%%CITATION = ARXIV:1611.02959;%%.

\bibitem{Ugajin:2016opf}
T.~Ugajin, ``{Mutual information of excited states and relative entropy of two
  disjoint subsystems in CFT},''
  \href{http://dx.doi.org/10.1007/JHEP10(2017)184}{{\em JHEP} {\bfseries 10}
  (2017) 184},
\href{http://arxiv.org/abs/1611.03163}{{\ttfamily arXiv:1611.03163 [hep-th]}}.
%%CITATION = ARXIV:1611.03163;%%.

\bibitem{Faulkner:2017tkh}
T.~Faulkner, F.~M. Haehl, E.~Hijano, O.~Parrikar, C.~Rabideau, and
  M.~Van~Raamsdonk, ``{Nonlinear Gravity from Entanglement in Conformal Field
  Theories},'' \href{http://dx.doi.org/10.1007/JHEP08(2017)057}{{\em JHEP}
  {\bfseries 08} (2017) 057},
\href{http://arxiv.org/abs/1705.03026}{{\ttfamily arXiv:1705.03026 [hep-th]}}.
%%CITATION = ARXIV:1705.03026;%%.

\bibitem{Sarosi:2017rsq}
G.~S\'arosi and T.~Ugajin, ``{Modular Hamiltonians of excited states, OPE
  blocks and emergent bulk fields},''
  \href{http://dx.doi.org/10.1007/JHEP01(2018)012}{{\em JHEP} {\bfseries 01}
  (2018) 012},
\href{http://arxiv.org/abs/1705.01486}{{\ttfamily arXiv:1705.01486 [hep-th]}}.
%%CITATION = ARXIV:1705.01486;%%.

\bibitem{Hollands:2017dov}
S.~Hollands and K.~Sanders, ``{Entanglement measures and their properties in
  quantum field theory},''
\href{http://arxiv.org/abs/1702.04924}{{\ttfamily arXiv:1702.04924
  [quant-ph]}}.
%%CITATION = ARXIV:1702.04924;%%.

\bibitem{Hollands:2017mlk}
S.~Hollands, O.~Islam, and K.~Sanders, ``{Relative entanglement entropy for
  widely separated regions in curved spacetime},''
\href{http://arxiv.org/abs/1711.02039}{{\ttfamily arXiv:1711.02039 [math-ph]}}.
%%CITATION = ARXIV:1711.02039;%%.

\bibitem{Nakaguchi:2014pha}
Y.~Nakaguchi and T.~Nishioka, ``{Entanglement Entropy of Annulus in Three
  Dimensions},'' \href{http://dx.doi.org/10.1007/JHEP04(2015)072}{{\em JHEP}
  {\bfseries 04} (2015) 072},
\href{http://arxiv.org/abs/1501.01293}{{\ttfamily arXiv:1501.01293 [hep-th]}}.
%%CITATION = ARXIV:1501.01293;%%.

\bibitem{Witten:2018zxz}
E.~Witten, ``{Notes on Some Entanglement Properties of Quantum Field Theory},''
\href{http://arxiv.org/abs/1803.04993}{{\ttfamily arXiv:1803.04993 [hep-th]}}.
%%CITATION = ARXIV:1803.04993;%%.

\bibitem{Duan2017}
Z.~Duan, L.~Niu, Y.~Wang, and L.~Liu, ``Relative entropy and relative entropy
  of entanglement for infinite-dimensional systems,''
  \href{http://dx.doi.org/10.1007/s10773-017-3338-2}{{\em International Journal
  of Theoretical Physics} {\bfseries 56} no.~6, (Jun, 2017) 1929--1936}.

\bibitem{DonaldHorodecki99:Continuity}
M.~J. Donald and M.~Horodecki, ``Continuity of relative entropy of
  entanglement,''
  \href{http://dx.doi.org/https://doi.org/10.1016/S0375-9601(99)00813-0}{{\em
  Physics Letters A} {\bfseries 264} no.~4, (1999) 257 -- 260}.

\bibitem{0305-4470-33-22-101}
M.~B. Plenio, S.~Virmani, and P.~Papadopoulos, ``Operator monotones, the
  reduction criterion and the relative entropy,'' {\em Journal of Physics A:
  Mathematical and General} {\bfseries 33} no.~22, (2000) L193.

\bibitem{Devetak207}
I.~Devetak and A.~Winter, ``Distillation of secret key and entanglement from
  quantum states,'' \href{http://dx.doi.org/10.1098/rspa.2004.1372}{{\em
  Proceedings of the Royal Society of London A: Mathematical, Physical and
  Engineering Sciences} {\bfseries 461} no.~2053, (2005) 207--235},
  \href{http://arxiv.org/abs/quant-ph/0306078}{{\ttfamily quant-ph/0306078}}.

\bibitem{1464-4266-6-12-009}
A.~Miranowicz and A.~Grudka, ``A comparative study of relative entropy of
  entanglement, concurrence and negativity,'' {\em Journal of Optics B: Quantum
  and Semiclassical Optics} {\bfseries 6} no.~12, (2004) 542,
  \href{http://arxiv.org/abs/quant-ph/0409153}{{\ttfamily quant-ph/0409153}}.

\bibitem{0305-4470-36-20-316}
J.~Eisert, K.~Audenaert, and M.~B. Plenio, ``Remarks on entanglement measures
  and non-local state distinguishability,'' {\em Journal of Physics A:
  Mathematical and General} {\bfseries 36} no.~20, (2003) 5605,
  \href{http://arxiv.org/abs/quant-ph/0212007}{{\ttfamily quant-ph/0212007}}.

\bibitem{Sep}
M.~{Lewenstein}, D.~{Bru{\ss}}, J.~I. {Cirac}, B.~{Kraus}, M.~{Kus},
  J.~{Samsonowicz}, A.~{Sanpera}, and R.~{Tarrach}, ``{Separability and
  distillability in composite quantum systems-a primer},''
  \href{http://dx.doi.org/10.1080/09500340008232176}{{\em Journal of Modern
  Optics} {\bfseries 47} (Nov., 2000) 2481--2499},
  \href{http://arxiv.org/abs/quant-ph/0006064}{{\ttfamily quant-ph/0006064}}.

\bibitem{SepV}
L.~{Gurvits} and H.~{Barnum}, ``{Largest separable balls around the maximally
  mixed bipartite quantum state},''
  \href{http://dx.doi.org/10.1103/PhysRevA.66.062311}{{\em Phys.Rev. A}
  {\bfseries 66} no.~6, (Dec., 2002) 062311},
  \href{http://arxiv.org/abs/quant-ph/0204159}{{\ttfamily quant-ph/0204159}}.

\bibitem{Ryu:2006bv}
S.~Ryu and T.~Takayanagi, ``{Holographic derivation of entanglement entropy
  from AdS/CFT},'' \href{http://dx.doi.org/10.1103/PhysRevLett.96.181602}{{\em
  Phys. Rev. Lett.} {\bfseries 96} (2006) 181602},
\href{http://arxiv.org/abs/hep-th/0603001}{{\ttfamily arXiv:hep-th/0603001
  [hep-th]}}.
%%CITATION = HEP-TH/0603001;%%.

\bibitem{Hayden:2011ag}
P.~Hayden, M.~Headrick, and A.~Maloney, ``{Holographic Mutual Information is
  Monogamous},'' \href{http://dx.doi.org/10.1103/PhysRevD.87.046003}{{\em Phys.
  Rev.} {\bfseries D87} no.~4, (2013) 046003},
\href{http://arxiv.org/abs/1107.2940}{{\ttfamily arXiv:1107.2940 [hep-th]}}.
%%CITATION = ARXIV:1107.2940;%%.

\bibitem{Calabrese:2009ez}
P.~Calabrese, J.~Cardy, and E.~Tonni, ``{Entanglement entropy of two disjoint
  intervals in conformal field theory},''
  \href{http://dx.doi.org/10.1088/1742-5468/2009/11/P11001}{{\em J. Stat.
  Mech.} {\bfseries 0911} (2009) P11001},
\href{http://arxiv.org/abs/0905.2069}{{\ttfamily arXiv:0905.2069 [hep-th]}}.
%%CITATION = ARXIV:0905.2069;%%.

\bibitem{Cardy:2013nua}
J.~Cardy, ``{Some results on the mutual information of disjoint regions in
  higher dimensions},''
  \href{http://dx.doi.org/10.1088/1751-8113/46/28/285402}{{\em J. Phys.}
  {\bfseries A46} (2013) 285402},
\href{http://arxiv.org/abs/1304.7985}{{\ttfamily arXiv:1304.7985 [hep-th]}}.
%%CITATION = ARXIV:1304.7985;%%.

\end{thebibliography}\endgroup

\end{document}